\documentclass[12pt,a4paper]{article}
\newcommand{\n}{\noindent}
\newcommand{\ket}[1]{\left|#1\right\rangle}      

\begin{document}

\newpage
\setcounter{page}{0}
\begin{titlepage}
\begin{flushright}
UFSCARF-TH-03-05
\end{flushright}
\vskip 2cm
\begin{center}
{\large  Spectrum of the supersymmetric t-J model \\
with non-diagonal open boundaries}\\
\vspace{1cm}
{\large W. Galleas } \\
\vspace{1cm}
{\em Universidade Federal de S\~ao Carlos\\
Departamento de F\'{\i}sica \\
C.P. 676, 13565-905, S\~ao Carlos-SP, Brasil}\\
\end{center}
\vspace{1.5cm}

\begin{abstract}
In this work we diagonalize the double-row transfer matrix of the supersymmetric t-J model with non-diagonal
boundary terms by means of the algebraic Bethe ansatz. The corresponding reflection equations are studied and
two distinct classes of solutions are found, one diagonal solution and other non-diagonal. In the non-diagonal case
the eigenvalues in the first sectors are given for arbitrary values of the boundary parameters.

\end{abstract}

\vspace{2.5cm}
\centerline{{\small PACS numbers:  05.50+q, 02.30.IK}}
\vspace{.1cm}
\centerline{{\small Keywords: Algebraic Bethe Ansatz, Lattice Models, Open Boundary Conditions}}
\vspace{2.0cm}
\centerline{{\small March 2007}}
\end{titlepage}

\section{Introduction}

The t-J model is frequently invoked as a model for strongly correlated electrons systems, in particular,
for high-$T_{c}$ cuprate superconductors \cite{ZR,SCH1} as well as for heavy-fermion systems \cite{SCH2,GROS}.
Although the mechanisms proposed to explain high-$T_{c}$ superconductors usually invoke properties
of two-dimensional systems \cite{ZR,KANE,AND}, it has been argued \cite{AND,AND1} that due to strong
quantum fluctuations one-dimensional systems may share some features of the two-dimensional ones.
The t-J hamiltonian contains nearst-neighbor hopping and nearst-neighbor spin exchange terms and it can
be derived from the Hubbard hamiltonian for a band occupation close to half-filling and a large
Coulomb repulsion by means of a canonical transformation which eliminates doubly occupied
states \cite{BRINK}. For a review on this subject see for instance the Ref. \cite{SCH3}.
The Hilbert space of this model is contrained to forbid double occupancy of single sites, leading to
only three possible states at each lattice site. At the supersymmetric point, the one-dimensional
t-J model becomes $sl(2|1)$ invariant and its integrability was first stated by Lai \cite{LAI}
and Sutherland \cite{SU}, and subsequently reported by other authors \cite{SCH2,SAR,BAR}. Within the Quantum Inverse
Scattering Method, the integrability of the supersymmetric t-J model was established in \cite{KOR} for the
case of periodic boundary conditions.

Though boundary conditions are not expected to influence
the infinite volume properties, it can modify the finite-size corrections of massless systems in a strip
of width $L$ which provides fundamental informations concerning the underlying
conformal field theories \cite{CARDY}. On the other hand, boundary conditions also provides a mechanism
to relate the critical behaviour of a variety of lattice systems such as the Heisenberg spin chain,
the Ashkin-Teller and Potts models \cite{ALCARAZ}.

The study of integrable systems with arbitrary boundary conditions gained a tremendous impulse with
Sklyanin's \cite{SK} generalization of the Quantum Inverse Scattering Method \cite{FAD} to accomodate
the case of open boundaries. In Sklyanin's approach, the construction of such models are based on solutions
of the so-called  reflection equations \cite{SK,CHER} for a given bulk system.
For the supersymmetric t-J model, two classes of diagonal solutions of the reflection equations were found
in \cite{GONZA} corresponding to boundary chemical potentials and boundary magnetic fields.
Physical properties like ground state structure and boundary susceptibilities have been
studied in \cite{ESSLER}.

However, general non-diagonal open boundaries for the supersymmetric t-J model have not been considered
so far in the literature. To our knowledge, the $Z_{N}$ Belavin model \cite{SASAKI}, the
$SU(N)$ vertex model \cite{WG} and the spin-$S$ Heisenberg chain \cite{CLA} are the only multistates
systems investigated so far with non-diagonal open boundaries. The recent progresses on this
matter are mostly concentrated on the eight \cite{FAN} and six \cite{NEPO,CAO,GIER} vertex
models when the boundary parameters satisfy certain constraints.

Non-diagonal solutions of the reflection equations are known for a variety of integrable models based on $q$-deformed Lie algebras
\cite{ABAD,LIMA,LM} but similar results concerning superalgebras, where the supersymmetric t-J model is inserted,
are still concentrated on the $U_{q}[osp(1|2)]$ symmetry \cite{ALS} and
on diagonal solutions \cite{GUANG0}. The main result of this paper
is to show that the covering transfer matrix of the supersymmetric t-J model built from a
general non-diagonal solution of the reflection equation possess a trivial reference state
needed to initiate an algebraic Bethe ansatz analysis.

This paper is organized as follows. In the next section we derive general solutions of the reflection equations
for the supersymmetric t-J model. Two classes of solutions were found, a non-diagonal one with five free
parameters and a diagonal solution with only one free parameter. In section 3 we perform the algebraic Bethe
ansatz analysis and conclusions and remarks are presented in section 4.

\section{Solutions of the Reflection Equation}

In Sklyanin's approach \cite{SK}, the first
step towards the construction of integrable models with open boundaries is to search for solutions
of the reflection equations for a given integrable bulk system. This equation governs the integrability
at the boundaries and it reads
\begin{equation}
\label{rem}
R_{21}(\lambda - \mu) K_{2}^{-}(\lambda) R_{12}(\lambda + \mu) K_{1}^{-}(\mu) =
K_{1}^{-}(\mu) R_{21}(\lambda + \mu) K_{2}^{-}(\lambda) R_{12}(\lambda - \mu).
\end{equation}
The matrix $K^{-}(\lambda)$ describes the reflection at one of the ends of an open chain, and a similar
equation should also hold for a matrix $K^{+}(\lambda)$ describing the reflection at the opposite boundary.
In its turn the $R$-matrix entering in the reflection equation (\ref{rem}) is a solution of the
Yang-Baxter equation, namely
\begin{equation}
\label{YB}
R_{12}(\lambda-\mu) R_{13}(\lambda) R_{23}(\mu) = R_{23}(\mu) R_{13}(\lambda) R_{12}(\lambda-\mu)
\end{equation}
defined in $V \otimes V \otimes V$. Here $V$ is a finite dimensional $Z_2$ graded space and
$R_{ij}(\lambda)$ consist of $R(\lambda) \in \mbox{End} [V \otimes V]$ acting non trivially 
in the $i$th and $j$th spaces of $V \otimes V \otimes V$.

The tensor products appearing in Eqs. (\ref{rem}) and (\ref{YB}) are defined as
$\left[ A \otimes B \right]^{ik}_{jl}
=A_{j}^{i} B_{l}^{k} (-1)^{(p_{i}+p_{j})p_{k}}$ for generic matrices $A$ and $B$.
This tensor product is equipped with Grassmann parities $p_{i}$ assuming values on the group $Z_2$ which enable us to distinguish bosonic
and fermionic degrees of freedom.

For the supersymmetric t-J model, the covering $R$-matrix consists of the rational $sl(2|1)$
invariant solution of the Yang-Baxter equation,
\begin{eqnarray}
\label{tj}
R(\lambda) = \sum_{i=1}^{3} a_{i}(\lambda) \hat{e}_{ii} \otimes \hat{e}_{ii}
+ b(\lambda) \sum_{i \neq j}^{3} (-1)^{p_{i} p_{j}} \hat{e}_{jj} \otimes \hat{e}_{ii}
+ \sum_{i \neq j}^{3} \hat{e}_{ji} \otimes \hat{e}_{ij},
\end{eqnarray}
\n where the Boltzmann weights are explicitly given by $a_{i}(\lambda)= 1+(-1)^{p_{i}} \lambda$
and $b(\lambda)= \lambda$. The elements $\hat{e}_{ij}$ denotes usual
$3 \times 3$  Weyl matrices with components
$\left( \hat{e}_{ij} \right)_{\alpha \beta} = \delta_{i \alpha} \delta_{j \beta}$ and in what follows we
have adopted the grading $p_{1}=1$ and $p_{2}=p_{3}=0$.

The $R$-matrix given in (\ref{tj}) satisfies important symmetries relations namely,
\begin{eqnarray}
&& \mbox{PT-symmetry} \;\;\;\;\;\; : \;\;\;\;\;\; P_{12} R_{12}(\lambda) P_{12} \;\; = R_{12}^{st_{1} st_{2}} (\lambda) \nonumber \\
&& \mbox{Unitarity} \;\;\;\;\;\;\;\;\;\;\;\;\; : \;\;\;\;\;\;  R_{12}(\lambda) R_{12}(-\lambda) = \left(1-\lambda^2 \right) \mbox{Id}\otimes \mbox{Id} \nonumber \\
&& \mbox{Cross-Unitarity} \;\;\; \; : \;\;\;\;\;\;  R_{12}^{st_1} (\lambda) M_{1}   R_{21}^{st_1}(-\lambda-2\rho) M_{1}^{-1} =
\zeta(\lambda) \mbox{Id}\otimes \mbox{Id}\nonumber
\end{eqnarray}
In the above relations Id is the $3\times 3$ identity matrix, $\zeta(\lambda)$ is a convenient normalization function and the matrix $M$
is a symmetry of the $R$-matrix, i.e.
\begin{equation}
\left[ R(\lambda) , M \otimes M \right]=0.
\end{equation}
\n The symbol $st_{k}$ stands for the supertransposition in the space labeled by the
index $k$ and $\displaystyle P_{12}=\sum_{i,j=1}^{3} (-1)^{p_{i} p_{j}} \hat{e}_{ij} \otimes \hat{e}_{ji}$
denotes the graded permutator.

When these properties are fulfilled one can follow the scheme devised in \cite{B11,B12}. In this
way the matrix $K^{-}(\lambda)$ is obtained by solving (\ref{rem}) and the matrix $K^{+}(\lambda)$ follows
from the isomorphism
\begin{equation}
\label{iso}
K^{-}(\lambda) \mapsto K^{+}(\lambda)^{st}=K^{-}(-\lambda-\rho) M.
\end{equation}
In the case considered here we have $\rho=\frac{1}{2}$ and the matrix
$M$ is the identity matrix.

Now we shall look for solutions of the reflection equation (\ref{rem}) in the most general form
\begin{equation}
\label{kgen}
K^{-(l)} (\lambda) = \sum_{i,j=1}^{3} k^{-(l)}_{ij}(\lambda) \hat{e}_{ij} .
\end{equation}
Substituting (\ref{kgen}) and the $R$-matrix (\ref{tj}) in (\ref{rem}) we are left with a system
of functional equations for the matrix elements $k^{-(l)}_{ij}(\lambda)$. A brute force
analysis of these equations allow us to identify two branches of solutions
\begin{eqnarray}
\label{sol}
K^{-(1)}(\lambda)=\pmatrix{
k^{-(1)}_{11}(\lambda) & 0 & 0 \cr
0 & k^{-(1)}_{22}(\lambda) & k^{-(1)}_{23}(\lambda) \cr
0 & k^{-(1)}_{32}(\lambda) & k^{-(1)}_{33}(\lambda) \cr} \;\;\;
K^{-(2)}(\lambda)=\pmatrix{
k^{-(2)}_{11}(\lambda) & 0 & 0 \cr
0 & k^{-(2)}_{22}(\lambda) & 0 \cr
0 & 0 & k^{-(2)}_{33}(\lambda) \cr}. \nonumber \\
\end{eqnarray}
The upper index $(l)$ is aimed to distinguish these two branches and their non-null elements are determined from
the following equations:
\begin{itemize}
\item Branch (1):
\end{itemize}
\begin{eqnarray}
\label{b1}
2 \lambda k^{-(1)}_{11}(\lambda) \frac{k^{-(1)}_{32}(\mu)}{k^{-(1)}_{32}(\lambda)} \left[ k^{-(1)}_{11}(\mu) - k^{-(1)}_{33}(\mu) \right] &=&
(\lambda+\mu) k^{-(1)}_{11}(\mu)^{2} - 2\lambda k^{-(1)}_{33}(\mu) k^{-(1)}_{11}(\mu) \nonumber \\
&+& (\lambda - \mu) k^{-(1)}_{33}(\mu)^{2} + (\lambda-\mu) k^{-(1)}_{23}(\mu) k^{-(1)}_{32}(\mu) \nonumber \\
\\
2 \lambda k^{-(1)}_{22}(\lambda) \frac{k^{-(1)}_{32}(\mu)}{k^{-(1)}_{32}(\lambda)} \left[ k^{-(1)}_{11}(\mu) - k^{-(1)}_{33}(\mu) \right] &=&
(\lambda+\mu) k^{-(1)}_{11}(\mu)^{2} - 2\lambda k^{-(1)}_{33}(\mu) k^{-(1)}_{11}(\mu) \nonumber \\
&+&(\lambda - \mu) k^{-(1)}_{33}(\mu)^{2} - (\lambda+\mu) k^{-(1)}_{23}(\mu) k^{-(1)}_{32}(\mu)  \nonumber \\
\\
2 \lambda k^{-(1)}_{33}(\lambda) \frac{k^{-(1)}_{32}(\mu)}{k^{-(1)}_{32}(\lambda)} \left[ k^{-(1)}_{11}(\mu) - k^{-(1)}_{33}(\mu) \right] &=&
- (\lambda-\mu) k^{-(1)}_{11}(\mu)^{2} + 2\lambda k^{-(1)}_{33}(\mu) k^{-(1)}_{11}(\mu) \nonumber \\
&-& (\lambda + \mu) k^{-(1)}_{33}(\mu)^{2} + (\lambda-\mu) k^{-(1)}_{23}(\mu) k^{-(1)}_{32}(\mu) \nonumber \\
\end{eqnarray}
\begin{eqnarray}
k^{-(1)}_{22}(\mu) \left[ k^{-(1)}_{11}(\mu) - k^{-(1)}_{33}(\mu) \right] &=& k^{-(1)}_{11}(\mu)^{2} - k^{-(1)}_{33}(\mu) k^{-(1)}_{11}(\mu) - k^{-(1)}_{23}(\mu) k^{-(1)}_{32}(\mu)  \\
\nonumber \\
\label{b1f}
k^{-(1)}_{23}(\lambda) k^{-(1)}_{32}(\mu) &=& k^{-(1)}_{23}(\mu) k^{-(1)}_{32}(\lambda)
\end{eqnarray}
\begin{itemize}
\item Branch (2):
\end{itemize}
\begin{eqnarray}
\label{b2}
k^{-(2)}_{11}(\lambda) \left[ (\lambda - \mu) k^{-(2)}_{11}(\mu) - (\lambda + \mu) k^{-(2)}_{22}(\mu) \right] &=&
-k^{-(2)}_{22}(\lambda) \left[ (\lambda + \mu) k^{-(2)}_{11}(\mu) + (\mu - \lambda) k^{-(2)}_{22}(\mu) \right] \nonumber \\
\\
\label{b2f}
k^{-(2)}_{22}(\lambda)&=&k^{-(2)}_{33}(\lambda)
\end{eqnarray}

The solution of Eqs. (\ref{b1}-\ref{b1f}) associated to the non-diagonal branch is given by
\begin{eqnarray}
\label{sol1}
k^{-(1)}_{11}(\lambda)&=&1 \nonumber \\
k^{-(1)}_{22}(\lambda)&=& \frac{h_{0}^{-} (h_{3}^{-} h_{4}^{-} - h_{1}^{-} h_{2}^{-}) + \lambda (h_{3}^{-}h_{4}^{-} + h_{1}^{-}h_{2}^{-})}{(h_{0}^{-}-\lambda)(h_{3}^{-}h_{4}^{-} - h_{1}^{-}h_{2}^{-})}
\;\;\;\;\;\;
k^{-(1)}_{23}(\lambda)=\frac{2\lambda h_{2}^{-} h_{3}^{-}}{(h_{0}^{-} - \lambda)(h_{3}^{-}h_{4}^{-} - h_{1}^{-}h_{2}^{-})} \nonumber \\
k^{-(1)}_{33}(\lambda)&=& \frac{h_{0}^{-} (h_{3}^{-}h_{4}^{-} - h_{1}^{-}h_{2}^{-}) - \lambda (h_{3}^{-}h_{4}^{-} + h_{1}^{-}h_{2}^{-})}{(h_{0}^{-}-\lambda)(h_{3}^{-}h_{4}^{-} - h_{1}^{-}h_{2}^{-})}
\;\;\;\;\;\;
k^{-(1)}_{32}(\lambda)= \frac{2\lambda h_{1}^{-} h_{4}^{-}}{(h_{0}^{-} - \lambda)(h_{1}^{-}h_{2}^{-} - h_{3}^{-}h_{4}^{-})}, \nonumber \\
\end{eqnarray}
\n while the solution of Eqs. (\ref{b2}-\ref{b2f}) turns out to be
\begin{eqnarray}
\label{sol2}
k^{-(2)}_{11}(\lambda) &=& \frac{h_0^{-} + \lambda}{h_0^{-} - \lambda} \nonumber \\
k^{-(2)}_{22}(\lambda) &=& k^{-(2)}_{33}(\lambda) = 1 .
\end{eqnarray}

The diagonal solution $K^{-(2)}(\lambda)$ possess only one free parameter $\{h^{-}_0 \}$
and it is contained in the rational limit of one of the solutions presented in \cite{GONZA} for the
$q$-deformed t-J model.
On the other hand, the solution $K^{- (1)}(\lambda)$
has altogether five free parameters $\{h^{-}_0 , h^{-}_1 , h^{-}_2 , h^{-}_3 , h^{-}_4 \}$ and it reduces 
to the second known diagonal solution by setting $h^{-}_3 = h^{-}_4 =0$. Moreover, the
non-diagonal solution $K^{- (1)}(\lambda)$ has null entries in suitable positions for an algebraic
Bethe ansatz study that will be explored in the next section. The question if this is the general
scenario for the $sl(m|n)$ symmetry deserves to be further studied.

Next we turn our attention to the matrices $K^{+ (l)}(\lambda)$ which follows immediately from the isomorphism
(\ref{iso}). Thus we have
\begin{eqnarray}
\label{psol}
K^{+(1)}(\lambda)=\pmatrix{
k^{+(1)}_{11}(\lambda) & 0 & 0 \cr
0 & k^{+(1)}_{22}(\lambda) & k^{+(1)}_{23}(\lambda) \cr
0 & k^{+(1)}_{32}(\lambda) & k^{+(1)}_{33}(\lambda) \cr}
\;\;\;
K^{+(2)}(\lambda)=\pmatrix{
k^{+(2)}_{11}(\lambda) & 0 & 0 \cr
0 & k^{+(2)}_{22}(\lambda) & 0 \cr
0 & 0 & k^{+(2)}_{33}(\lambda) \cr} \nonumber \\
\end{eqnarray}
\n where
\begin{eqnarray}
\label{psol1}
k^{+(1)}_{11}(\lambda)&=&1 \nonumber \\
k^{+(1)}_{22}(\lambda)&=& \frac{h_{0}^{+} (h_{3}^{+} h_{4}^{+} - h_{1}^{+} h_{2}^{+}) - (\lambda +\frac{1}{2}) (h_{3}^{+}h_{4}^{+} + h_{1}^{+}h_{2}^{+})}{(h_{0}^{+}+\frac{1}{2}+\lambda)(h_{3}^{+}h_{4}^{+} - h_{1}^{+}h_{2}^{+})}
\nonumber \\
k^{+(1)}_{33}(\lambda)&=& \frac{h_{0}^{+} (h_{3}^{+}h_{4}^{+} - h_{1}^{+}h_{2}^{+}) + (\lambda +\frac{1}{2})(h_{3}^{+}h_{4}^{+} + h_{1}^{+}h_{2}^{+})}{(h_{0}^{+}+\frac{1}{2}+\lambda)(h_{3}^{+}h_{4}^{+} - h_{1}^{+}h_{2}^{+})}
\nonumber \\
k^{+(1)}_{23}(\lambda)&=&\frac{2(\lambda+\frac{1}{2}) h_{2}^{+} h_{3}^{+}}{(h_{0}^{+} +\frac{1}{2}+ \lambda)(h_{3}^{+}h_{4}^{+} - h_{1}^{+}h_{2}^{+})} \nonumber \\
k^{+(1)}_{32}(\lambda)&=&\frac{2(\lambda+\frac{1}{2}) h_{1}^{+} h_{4}^{+}}{(h_{0}^{+} +\frac{1}{2}+ \lambda)(h_{1}^{+}h_{2}^{+} - h_{3}^{+}h_{4}^{+})},
\end{eqnarray}
\n and
\begin{eqnarray}
\label{psol2}
k^{+(2)}_{11}(\lambda) &=& \frac{h_0^{+} -\frac{1}{2}- \lambda}{h_0^{+} +\frac{1}{2}+ \lambda} \nonumber \\
k^{+(2)}_{22}(\lambda) &=& k^{+(2)}_{33}(\lambda) = 1 .
\end{eqnarray}

Here we remark that the matrices $K^{-(l)}(\lambda)$ defined by Eqs. (\ref{sol},\ref{sol1},\ref{sol2})
consist of regular solutions, i.e. $K^{-(l)}(0)=I$.
One can also verify that
\begin{equation}
\left[ K^{\pm (l)}(\lambda) , K^{\pm (l)}(\mu) \right]=0 ,
\end{equation}
\n which implies that the matrices $K^{\pm (l)}(\lambda)$ can be made diagonal by a similarity
transformation independent of the spectral parameter $\lambda$. This later property has been used
in \cite{WG,CLA}.

Now we have the basic ingredients to build an integrable model with open boundary conditions. 
Following \cite{SK}, the covering transfer matrix of the supersymmetric t-J model with open boundaries can
be written as the following supertrace over the $3 \times 3$ auxiliary space $\mathcal{A}$,

\begin{equation}
\label{tab}
T^{(l,m)}(\lambda)=\mbox{str}_{\mathcal{A}} \left[ K^{+(l)}_{\mathcal{A}}(\lambda) \mathcal{T}_{\mathcal{A}} (\lambda)
K^{-(m)}_{\mathcal{A}}(\lambda) \tilde{\mathcal{T}}_{\mathcal{A}} (\lambda) \right]   \;\;\;\;\;\;\;\;\;\;\;\;\;\;\;\; l,m=1,2
\end{equation}

\n where $\mathcal{T}_{\mathcal{A}} (\lambda)=R_{\mathcal{A} L}(\lambda) R_{\mathcal{A} L-1}(\lambda) \dots R_{\mathcal{A} 1}(\lambda)$
and $\tilde{\mathcal{T}}_{\mathcal{A}} (\lambda)=R_{\mathcal{A} 1}(\lambda) R_{\mathcal{A} 2}(\lambda) \dots R_{\mathcal{A} L}(\lambda)$
are the standard monodromy matrices that generate the corresponding closed t-J model
with $L$ sites \cite{KOR,FAD}.
We have also introduced the label $(l,m)$ in order to distinguish the $K$-matrices we are considering since
we can take $K^{\pm (l)}(\lambda)$ either from $K^{\pm (1)}(\lambda)$ or $K^{\pm (2)}(\lambda)$. Thus
altogether we have four different double-row transfer matrices.

Associated to each transfer matrix $T^{(l,m)}(\lambda)$ we have an
hamiltonian $\mathcal{H}^{(l,m)}$ with open boundary conditions given by
\begin{eqnarray}
\label{HAB}
\mathcal{H}^{(l,m)} &=& \sum_{j=1}^{L-1} \sum_{\sigma=\pm}
c^{\dagger}_{j,\sigma} \left( 1 - n_{j,-\sigma} \right) c_{j+1,\sigma} \left( 1 - n_{j+1,-\sigma} \right)
+ c^{\dagger}_{j+1,\sigma} \left( 1 - n_{j+1,-\sigma} \right) c_{j,\sigma} \left( 1 - n_{j,-\sigma} \right) \nonumber \\
&+& 2\sum_{j=1}^{L-1} \left[ S^{z}_{j} S^{z}_{j+1} + \frac{1}{2}\left(S^{+}_{j} S^{-}_{j+1}
+ S^{-}_{j} S^{+}_{j+1} \right) - \frac{1}{4} n_{j} n_{j+1} \right]
+ \sum_{j=1}^{L-1} \left( n_{j} + n_{j+1} \right) +  \mathcal{H}^{(l,m)}_{B} \nonumber \\
\end{eqnarray}
where the integrable boundary terms $\mathcal{H}^{(l,m)}_{B}$ are
\begin{eqnarray}
\label{BHAB}
\mathcal{H}^{(1,1)}_{B} &=& -\frac{(h^{-}_{1}h^{-}_{2}+h^{-}_{3}h^{-}_{4})}{h^{-}_{0}(h^{-}_{1}h^{-}_{2}-h^{-}_{3}h^{-}_{4})} S^{z}_{1}
-\frac{h^{-}_{2}h^{-}_{3}}{h^{-}_{0}(h^{-}_{1}h^{-}_{2}-h^{-}_{3}h^{-}_{4})} S^{+}_{1}
+\frac{h^{-}_{1}h^{-}_{4}}{h^{-}_{0}(h^{-}_{1}h^{-}_{2}-h^{-}_{3}h^{-}_{4})} S^{-}_{1}
+\frac{1}{2h^{-}_{0}} n_{1} \nonumber \\
&+&\frac{(h^{+}_{1}h^{+}_{2}+h^{+}_{3}h^{+}_{4})}{(h^{+}_{0} - \frac{1}{2})(h^{+}_{1}h^{+}_{2}-h^{+}_{3}h^{+}_{4})} S^{z}_{L}
+\frac{h^{+}_{2}h^{+}_{3}}{(h^{+}_{0} - \frac{1}{2})(h^{+}_{1}h^{+}_{2}-h^{+}_{3}h^{+}_{4})} S^{+}_{L} \nonumber \\
&-&\frac{h^{+}_{1}h^{+}_{4}}{(h^{+}_{0} - \frac{1}{2})(h^{+}_{1}h^{+}_{2}-h^{+}_{3}h^{+}_{4})} S^{-}_{L}
-\frac{1}{2(h^{+}_{0} - \frac{1}{2})} n_{L} \\
\mathcal{H}^{(1,2)}_{B} &=& -\frac{1}{h^{-}_{0}} n_{1} + \frac{(h^{+}_{1}h^{+}_{2}+h^{+}_{3}h^{+}_{4})}{(h^{+}_{0} - \frac{1}{2})(h^{+}_{1}h^{+}_{2}-h^{+}_{3}h^{+}_{4})} S^{z}_{L}
+\frac{h^{+}_{2}h^{+}_{3}}{(h^{+}_{0} - \frac{1}{2})(h^{+}_{1}h^{+}_{2}-h^{+}_{3}h^{+}_{4})} S^{+}_{L} \nonumber \\
&-&\frac{h^{+}_{1}h^{+}_{4}}{(h^{+}_{0} - \frac{1}{2})(h^{+}_{1}h^{+}_{2}-h^{+}_{3}h^{+}_{4})} S^{-}_{L}
-\frac{1}{2(h^{+}_{0} - \frac{1}{2})} n_{L} \\
\mathcal{H}^{(2,1)}_{B} &=& -\frac{(h^{-}_{1}h^{-}_{2}+h^{-}_{3}h^{-}_{4})}{h^{-}_{0}(h^{-}_{1}h^{-}_{2}-h^{-}_{3}h^{-}_{4})} S^{z}_{1}
-\frac{h^{-}_{2}h^{-}_{3}}{h^{-}_{0}(h^{-}_{1}h^{-}_{2}-h^{-}_{3}h^{-}_{4})} S^{+}_{1}
+\frac{h^{-}_{1}h^{-}_{4}}{h^{-}_{0}(h^{-}_{1}h^{-}_{2}-h^{-}_{3}h^{-}_{4})} S^{-}_{1}
+\frac{1}{2h^{-}_{0}} n_{1} \nonumber \\
&+& \frac{1}{(h^{+}_{0} + \frac{3}{2})} n_{L} \\
\mathcal{H}^{(2,2)}_{B} &=& -\frac{1}{h^{-}_{0}} n_{1} + \frac{1}{(h^{+}_{0} + \frac{3}{2})} n_{L}.
\label{BHABf}
\end{eqnarray}

The hamiltonians $\mathcal{H}^{(l,m)}$ are identified with the supersymmetric t-J model with open boundary conditions and,
omitting terms proportional to the identity, they are proportional to $\frac{d}{d\lambda} T^{(l,m)}(\lambda) \mid_{\lambda=0}$.
We have expressed the hamiltonians (\ref{HAB}-\ref{BHABf}) in terms of fermionic creation and annihilation
operators $c_{j,\sigma}^{\dagger}$ and $c_{j,\sigma}$ acting on the site $j$ and carrying spin
index $\sigma=\pm$. The spin operators $S^{+}_{j}$, $S^{-}_{j}$ and $S^{z}_{j}$ form an $su(2)$ algebra, and
the number operators are denoted $n_{j}=n_{j,+} + n_{j,-}$ where $n_{j,\sigma}=c^{\dagger}_{j,\sigma}c_{j,\sigma}$.
Except for the case $\mathcal{H}^{(2,2)}$, the remaining hamiltonians contain non-diagonal boundary
terms. In the next section we will discuss the diagonalization of the transfer matrices $T^{(l,m)}(\lambda)$
through the algebraic Bethe ansatz.

\section{Algebraic Bethe Ansatz}

The purpose of this section is to determine the eigenvalues and eigenvectors of the transfer matrices
defined in Eq. (\ref{tab}) built from the two branches of solution of the reflection equations.
For instance, when the $K$-matrices are diagonal this problem have been solved by the nested algebraic
Bethe ansatz \cite{GONZA}. Despite of the recent efforts in solving commuting transfer
matrices with general open boundary conditions \cite{SASAKI}-\cite{CAO}, the progresses are modest when compared with
the literature known for the diagonal case \cite{GUANG0,GUANG1}. Usually the diagonalization of double-row
transfer matrices with non-diagonal $K$-matrices is a tantalizing problem due to the difficulty in finding
a suitable reference state to perform a Bethe ansatz analysis. Here we note that the non-diagonal
$K$-matrices for the supersymmetric t-J model permit us to use the usual ferromagnetic state as
pseudovacuum state
\footnote{The author thanks M.J. Martins for pointing out this possibility.}.

In order to show that, we first represent the double-row monodromy matrix,
$U^{(l)}_{\mathcal{A}}(\lambda)= \mathcal{T}_{\mathcal{A}} (\lambda)
K^{-(l)}_{\mathcal{A}}(\lambda) \tilde{\mathcal{T}}_{\mathcal{A}} (\lambda)$, conveniently in the form
\begin{eqnarray}
\label{dmono}
U^{(l)}_{\mathcal{A}}(\lambda)= \pmatrix{
A^{(l)}(\lambda) & B^{(l)}_{1}(\lambda) & B^{(l)}_{2}(\lambda) \cr
C^{(l)}_{1}(\lambda) & D^{(l)}_{11}(\lambda) & D^{(l)}_{12}(\lambda) \cr
C^{(l)}_{2}(\lambda) & D^{(l)}_{21}(\lambda) & D^{(l)}_{22}(\lambda) \cr}.
\end{eqnarray}
Then the eigenvalue problem for the transfer matrix,
\begin{equation}
\label{eigp}
T^{(l,m)}(\lambda) \ket{\Psi} = \Lambda^{(l,m)} (\lambda) \ket{\Psi},
\end{equation}
\n becomes equivalent to
\begin{equation}
\label{eigpa}
\left[ -k^{+(l)}_{11}(\lambda) A^{(m)}(\lambda) + \sum_{i,j=1}^{2} k^{+(l)}_{i+1,j+1}(\lambda) D^{(m)}_{ji}(\lambda) \right]
\ket{\Psi} = \Lambda^{(l,m)} (\lambda) \ket{\Psi} .
\end{equation}

Next we consider the action of the elements of $U^{(l)}_{\mathcal{A}}(\lambda)$ on the
pseudovacuum state $\ket{\Psi_0}$ defined as
\begin{equation}
\label{psi0}
\ket{\Psi_{0}} = \bigotimes_{j=1}^{L} \ket{0}_{j} \;\;\;\;\;\;\;
\ket{0}_{j}=\pmatrix{
1 \cr 0 \cr 0 \cr}.
\end{equation}
The definitions (\ref{tab},\ref{dmono},\ref{psi0}) allow us to show that the elements of
$U^{(l)}_{\mathcal{A}}(\lambda)$ satisfy the relations
\begin{eqnarray}
\label{act}
A^{(l)} (\lambda) \ket{\Psi_0} &=& k^{-(l)}_{11}(\lambda) a_{1}^{2L}(\lambda) \ket{\Psi_0} \nonumber \\
D^{(l)}_{11} (\lambda) \ket{\Psi_0} &=& \left \{ \frac{k^{-(l)}_{11}(\lambda)}{a_{1}(2\lambda)} a_{1}^{2L}(\lambda) + \left[ k^{-(l)}_{22}(\lambda) -  \frac{k^{-(l)}_{11}(\lambda)}{a_{1}(2\lambda)} \right] b^{2L}(\lambda) \right\} \ket{\Psi_0} \nonumber \\
D^{(l)}_{22} (\lambda) \ket{\Psi_0} &=& \left \{ \frac{k^{-(l)}_{11}(\lambda)}{a_{1}(2\lambda)} a_{1}^{2L}(\lambda) + \left[ k^{-(l)}_{33}(\lambda) -  \frac{k^{-(l)}_{11}(\lambda)}{a_{1}(2\lambda)} \right] b^{2L}(\lambda) \right\} \ket{\Psi_0} \nonumber \\
D^{(l)}_{12} (\lambda) \ket{\Psi_0} &=& k^{-(l)}_{23}(\lambda)  b^{2L}(\lambda) \ket{\Psi_0}
\;\;\;\;\;\;\;\;\;\;\;\;\;\;\;\;
D^{(l)}_{21} (\lambda) \ket{\Psi_0} = k^{-(l)}_{32}(\lambda)  b^{2L}(\lambda) \ket{\Psi_0} \nonumber \\
B^{(l)}_{i} (\lambda) \ket{\Psi_0} &=& \dagger
\;\;\;\;\;\;\;\;\;\;\;\;\;\;\;\;\;\;\;\;\;\;\;\;\;\;\;\;\;\;\;\;\;\;\;\;\;\;\;\;\;\;\;\;
C^{(l)}_{i} (\lambda) \ket{\Psi_0} = 0
\;\;\;\;\;\;\;\;\;\;\;\;\;\; i=1,2
\end{eqnarray}
where the symbol $\dagger$ stands for a non-null value.

The relations (\ref{act}) together with the Eq. (\ref{eigpa})
imply that $\ket{\Psi_{0}}$ is an eigenvector of $T^{(l,m)}(\lambda)$ whose respective eigenvalue is
\begin{eqnarray}
\label{gama0}
\Lambda_{0}^{(l,m)}(\lambda) &=& \left\{ -k^{+(l)}_{11}(\lambda) k^{-(m)}_{11}(\lambda)
+\frac{k^{-(m)}_{11}(\lambda)}{a_{1}(2\lambda)} \sum_{i=2}^{3} k^{+(l)}_{ii}(\lambda) \right\} a_{1}^{2L}(\lambda) \nonumber \\
&+& \left\{ \sum_{i,j=2}^{3} k^{+(l)}_{ij}(\lambda) k^{-(m)}_{ji}(\lambda)
- \frac{k^{-(m)}_{11}(\lambda)}{a_{1}(2\lambda)}  \sum_{i=2}^{3} k^{+(l)}_{ii}(\lambda)  \right\}
b^{2L}(\lambda).
\end{eqnarray}

In the framework of the algebraic Bethe ansatz we now seek for other eigenvectors of $T^{(l,m)}(\lambda)$
in the multiparticle state form
\begin{equation}
\label{psi}
\ket{\Psi} = B^{(l)}_{a_{1}}(\lambda^{(1)}_{1}) B^{(l)}_{a_{2}}(\lambda^{(1)}_{2}) \dots B^{(l)}_{a_{n_{1}}}(\lambda^{(1)}_{n_{1}})
\mathcal{F}^{a_{1} a_{2} \dots a_{n_{1}}} \ket{\Psi_0}.
\end{equation}
In order to accomplish this task, the next step is to write appropriate commutation relations for
the elements of $U^{(l)}_{\mathcal{A}}(\lambda)$ which also satisfies
the quadratic relation (\ref{rem}) with $K^{-}(\lambda)$ replaced by $U^{(l)}_{\mathcal{A}}(\lambda)$.
From Eqs. (\ref{rem}) and (\ref{dmono}) it follows that
three of those commutation relations are of great use, namely
\begin{eqnarray}
\label{algSK}
A^{(l)}(\lambda) B^{(l)}_{j}(\mu) &=& \frac{a_{1}(\mu - \lambda)}{b(\mu - \lambda)} \frac{b(\mu + \lambda)}{a_{1}(\mu + \lambda)}
B^{(l)}_{j}(\mu) A^{(l)}(\lambda) - \frac{b(2\mu )}{a_{1}(2\mu )} \frac{1}{b(\mu - \lambda)} B^{(l)}_{j}(\lambda) A^{(l)}(\mu) \nonumber \\
&-&\frac{1}{a_{1}(\lambda + \mu)} B^{(l)}_{i}(\lambda) \tilde{D}^{(l)}_{ij}(\mu) \\
\label{algSK1}
\tilde{D}^{(l)}_{ij} (\lambda) B^{(l)}_{k}(\mu) &=& \frac{r^{id}_{ef} (\lambda +\mu -1) r^{fg}_{kj} (\lambda -\mu)}{b(\lambda+\mu-1) b(\lambda - \mu)}
B^{(l)}_{d}(\mu) \tilde{D}^{(l)}_{eg} (\lambda) +\frac{r^{id}_{ej} (2\lambda -1)}{a_{1}(2\lambda) b(\lambda - \mu)} B^{(l)}_{d}(\lambda) \tilde{D}^{(l)}_{ek}(\mu) \nonumber \\
&-& \frac{b(2\mu )}{a_{1}(2\mu )}  \frac{r^{id}_{kj} (2\lambda -1)}{a_{1}(2\lambda) a_{1}(\lambda + \mu)} B^{(l)}_{d}(\lambda) A^{(l)}(\mu) \\
B^{(l)}_{i}(\lambda) B^{(l)}_{j}(\mu) &=& B^{(l)}_{k}(\mu) B^{(l)}_{l}(\lambda) \frac{r^{ij}_{lk}(\lambda - \mu)}{a_{1}(\lambda - \mu)}  .
\end{eqnarray}

In the above relations we have defined the operator
\begin{equation}
\tilde{D}^{(l)}_{ij}(\lambda)= D^{(l)}_{ij}(\lambda) -\frac{\delta_{ij}}{a_{1}(2\lambda)} A^{(l)}(\lambda)
\end{equation}
\n and $r^{ik}_{jl} (\lambda)$ denotes the matrix elements of the rational $sl(2)$ $R$-matrix
\footnote{We have used the convention $\displaystyle r(\lambda)=\sum_{i,j,k,l}^{2} r^{ik}_{jl} (\lambda) \hat{e}_{ij}
\otimes \hat{e}_{kl}$, where now $\hat{e}_{ij}$ denotes $2\times 2$ Weyl matrices.}
\begin{eqnarray}
\label{sl2}
r(\lambda) = \pmatrix{
1+\lambda & 0 & 0 & 0 \cr
0 & \lambda & 1 & 0 \cr
0 & 1 & \lambda & 0 \cr
0 & 0 & 0 & 1+\lambda \cr}.
\end{eqnarray}

By carrying on the fields $A^{(l)}(\lambda)$ and  $D^{(l)}_{ij}(\lambda)$ over the
multiparticle state (\ref{psi}) we generate terms that are proportional to $\ket{\Psi}$
and others that are not. The terms proportional to $\ket{\Psi}$ contribute to the
eigenvalue while the other terms are usually denominated unwanted terms. The eigenvalue
$\Lambda^{(l,m)}(\lambda)$ is obtained from the first terms of the commutation relations
(\ref{algSK},\ref{algSK1}), together with the requirement that $\mathcal{F}^{a_{1} a_{2} \dots a_{n_{1}}}$ are components of
the eigenstates of an auxiliary double-row operator $\bar{T}^{(l,m)}(\lambda,\{ \lambda^{(1)}_{j} \})$ whose eigenvalue
equation reads
\begin{equation}
\bar{T}^{(l,m)}(\lambda,\{ \lambda^{(1)}_{j} \}) \ket{\mathcal{F}} =
\bar{\Lambda}^{(l,m)}(\lambda,\{ \lambda^{(1)}_{j} \}) \ket{\mathcal{F}} .
\end{equation}
This auxiliary operator is given by the following trace over the $2\times 2$
auxiliary space $\bar{\mathcal{A}}$
\begin{eqnarray}
\label{tnested}
&&\bar{T}^{(l,m)}(\lambda,\{ \lambda^{(1)}_{j} \}) = \nonumber \\
&& \mbox{Tr}_{\bar{\mathcal{A}}}
\left[ \bar{K}^{+(l)}_{\bar{\mathcal{A}}} (\lambda) r_{\bar{\mathcal{A}} 1}(\lambda+\lambda_{1}^{(1)}-1)
\dots r_{\bar{\mathcal{A}} n_{1}}(\lambda+\lambda_{n_{1}}^{(1)}-1)
\bar{K}^{-(m)}_{\bar{\mathcal{A}}} (\lambda) r_{\bar{\mathcal{A}} n_{1}}(\lambda-\lambda_{n_{1}}^{(1)})
\dots r_{\bar{\mathcal{A}} 1}(\lambda-\lambda_{1}^{(1)}) \right] \nonumber \\
\end{eqnarray}
and the associated $K$-matrices are given by
\begin{eqnarray}
\label{knest}
\bar{K}^{+(l)}(\lambda)=\pmatrix{
k^{+(l)}_{22}(\lambda) & k^{+(l)}_{23}(\lambda) \cr
k^{+(l)}_{32}(\lambda) & k^{+(l)}_{33}(\lambda) \cr}
\;\;\;\;\;\;\;\;\;\;
\bar{K}^{-(l)}(\lambda)=\pmatrix{
k^{-(l)}_{22}(\lambda) - \frac{k^{-(l)}_{11}(\lambda)}{a_{1}(2\lambda)} & k^{-(l)}_{23}(\lambda) \cr
k^{-(l)}_{32}(\lambda) & k^{-(l)}_{33}(\lambda) - \frac{k^{-(l)}_{11}(\lambda)}{a_{1}(2\lambda)} \cr}. \nonumber \\
\end{eqnarray}

For instance, the unwanted terms are eliminated provided the set of rapidities
$\{ \lambda^{(1)}_{j} \}$ satisfy the Bethe ansatz equations
\begin{eqnarray}
\label{eba}
&&\left[ \frac{a_{1}(\lambda_{j}^{(1)})}{b(\lambda_{j}^{(1)})} \right]^{2L}
\left[ \frac{1}{a_{1}(2\lambda^{(1)}_{j})} \sum_{i=2}^{3} k^{+(l)}_{ii}(\lambda^{(1)}_{j})
- k^{+(l)}_{11}(\lambda^{(1)}_{j}) \right] k^{-(m)}_{11}(\lambda^{(1)}_{j}) b(2\lambda^{(1)}_{j}) \nonumber \\
&&=(-1)^{n_{1}} \prod_{i \neq j}^{n_{1}} \frac{a_{1}(\lambda^{(1)}_{i}+\lambda^{(1)}_{j})}{b(\lambda^{(1)}_{i}+\lambda^{(1)}_{j})
a_{1}(\lambda^{(1)}_{i}-\lambda^{(1)}_{j}) b(\lambda^{(1)}_{j}+\lambda^{(1)}_{i}-1)}
\bar{\Lambda}^{(l,m)}(\lambda=\lambda^{(1)}_{j} ,\{ \lambda^{(1)}_{j} \}).
\end{eqnarray}

Thus we are left with the following expression for the eigenvalues $\Lambda^{(l,m)}(\lambda)$,
\begin{eqnarray}
\label{eigs}
\Lambda^{(l,m)}(\lambda) &=& \left[ \frac{1}{a_{1}(2\lambda)} \sum_{i=2}^{3} k^{+(l)}_{ii}(\lambda)
- k^{+(l)}_{11}(\lambda) \right] k^{-(m)}_{11}(\lambda)
a_{1}(\lambda)^{2L} \prod_{i=1}^{n_{1}}
\frac{a_{1}(\lambda^{(1)}_{i} - \lambda)}{b(\lambda^{(1)}_{i} - \lambda)} \frac{b(\lambda^{(1)}_{i} + \lambda)}{a_{1}(\lambda^{(1)}_{i} + \lambda)} \nonumber \\
&+& b(\lambda)^{2L} \prod_{i=1}^{n_{1}}
\frac{1}{b(\lambda-\lambda^{(1)}_{i}) b(\lambda+\lambda^{(1)}_{i}-1)}
\bar{\Lambda}^{(l,m)}(\lambda ,\{ \lambda^{(1)}_{j} \}).
\end{eqnarray}
This completes only the first step of the Bethe ansatz analysis since we still need to determine the
eigenvalues $\bar{\Lambda}^{(l,m)}(\lambda ,\{ \lambda^{(1)}_{j} \})$.

The auxiliary transfer matrix $\bar{T}^{(l,m)}(\lambda ,\{ \lambda^{(1)}_{j} \})$ defined in Eq. (\ref{tnested})
corresponds to that of an inhomogeneous $sl(2)$ vertex model with non-diagonal open boundaries.
Tipically, this problem is carried out by a second Bethe ansatz.
Here we note that the eigenvalues of $\bar{T}^{(l,m)}(\lambda ,\{ \lambda^{(1)}_{j} \})$ for
the case $n_{1}=1$ can be obtained by conventional methods.

For the case $n_{1}=1$ the auxiliary double-row operator reads
\begin{equation}
\label{nest1}
\bar{T}^{(l,m)}(\lambda,\{ \lambda^{(1)}_{1} \}) = \mbox{Tr}_{\bar{\mathcal{A}}}
\left[ \bar{K}^{+(l)}_{\bar{\mathcal{A}}} (\lambda) r_{\bar{\mathcal{A}} 1}(\lambda+\lambda_{1}^{(1)}-1)
\bar{K}^{-(m)}_{\bar{\mathcal{A}}} (\lambda)
r_{\bar{\mathcal{A}} 1}(\lambda-\lambda_{1}^{(1)}) \right],
\end{equation}
\n which consist of a $2 \times 2$ matrix. The secular equation, $\mbox{det} \left[ \bar{T}^{(l,m)}(\lambda,\{ \lambda^{(1)}_{1} \}) - \bar{\Lambda}^{(l,m)}(\lambda,\{ \lambda^{(1)}_{1} \})  \right] =0$,
in this case gives 
\begin{equation}
\label{sec1}
X^{(l,m)}_{2} \bar{\Lambda}^{(l,m)}(\lambda,\{ \lambda^{(1)}_{1} \})^2 +
X^{(l,m)}_{1} \bar{\Lambda}^{(l,m)}(\lambda,\{ \lambda^{(1)}_{1} \}) +
X^{(l,m)}_{0} =0,
\end{equation}
leaving us with the following expression for the eigenvalues 
$\bar{\Lambda}^{(l,m)}(\lambda,\{ \lambda^{(1)}_{1} \})$,
\begin{eqnarray}
\label{soln1}
\bar{\Lambda}^{(l,m)}(\lambda,\{ \lambda^{(1)}_{1} \}) = \frac{-X^{(l,m)}_{1} \pm \sqrt{(X^{(l,m)}_{1})^2 - 4X^{(l,m)}_{2} X^{(l,m)}_{0}}}{2 X^{(l,m)}_{2}}.
\end{eqnarray}
In order to avoid an overcrowded section we have collected the functions $X^{(l,m)}_{i}$
in the appendix A.
The relation (\ref{soln1}) together with (\ref{eba}) and (\ref{eigs}) determines the 
eigenvalues of $T^{(l,m)}(\lambda)$ in the sector $n_{1}=1$ without any constraint for the boundary parameters.

For general $n_{1}$ the nested problem consists in the diagonalization of the auxiliary transfer matrix
$\bar{T}^{(l,m)}(\lambda,\{ \lambda^{(1)}_{j} \})$ whose dimension is $2^{n_{1}} \times 2^{n_{1}}$.
This problem can be tackled by a second Bethe ansatz in the lines of \cite{WG,CLA}. Except for the case
$\bar{T}^{(1,1)}(\lambda,\{ \lambda^{(1)}_{j} \})$, no constraint is required for the boundary parameters. 
 
In order to do that we
first note that the associated boundary matrices $\bar{K}^{\pm(l)}(\lambda)$ also satisfy
$\left[ \bar{K}^{\pm(l)}(\lambda), \bar{K}^{\pm(l)}(\mu) \right]=0$. Thus they can be written as
\begin{eqnarray}
\label{ndec}
\bar{K}^{\pm(l)}(\lambda)=\mathcal{G}^{\pm(l)} D^{\pm(l)}(\lambda) \left( \mathcal{G}^{\pm(l)} \right)^{-1}
\end{eqnarray}
where $D^{\pm(l)}(\lambda)$ is a diagonal matrix and $\mathcal{G}^{\pm(l)}$ is independent of the
spectral parameter $\lambda$.
The matrices $\mathcal{G}^{\pm(1)}$ are then given by
\begin{eqnarray}
\label{M}
\mathcal{G}^{\pm(1)} = \pmatrix{
h^{\pm}_{2} & - h^{\pm}_{3} \cr
-h^{\pm}_{4} & h^{\pm}_{1} \cr},
\end{eqnarray}
while the diagonal matrices $D^{\pm(1)}(\lambda)$ turn out to be
\begin{eqnarray}
\label{DiagK}
D^{+(1)}(\lambda)=\pmatrix{
1 & 0 \cr
0 & \frac{h^{+}_{0} - \lambda - \frac{1}{2}}{h^{+}_{0} + \lambda + \frac{1}{2}} \cr} \;\;\;\; \;\;\;\;
D^{-(1)}(\lambda)=\frac{\lambda}{\lambda - \frac{1}{2}} \pmatrix{
1 & 0 \cr
0 & \frac{h^{-}_{0} + \lambda - 1}{h^{-}_{0} - \lambda} \cr}.
\end{eqnarray}
By way of contrast, the matrices $\bar{K}^{\pm(2)}(\lambda)$ are proportional to the identity and
the relation (\ref{ndec}) becomes trivial.

Next we proceed by inserting terms $\mathcal{G}^{+(l)}_{\bar{\mathcal{A}}} \left( \mathcal{G}^{+(l)}_{\bar{\mathcal{A}}} \right)^{-1}$
in between all the elements of the double-row operator (\ref{tnested}). In this way the transfer matrix
$\bar{T}^{(l,m)}(\lambda,\{ \lambda^{(1)}_{j} \})$ can be rewritten as
\begin{eqnarray}
\label{Ntnest}
&&\bar{T}^{(l,m)}(\lambda,\{ \lambda^{(1)}_{j} \}) = \nonumber \\
&& \mbox{Tr}_{\bar{\mathcal{A}}}
\left[ D^{+(l)}_{\bar{\mathcal{A}}} (\lambda) \tilde{r}^{(l)}_{\bar{\mathcal{A}} 1}(\lambda+\lambda_{1}^{(1)}-1)
\dots \tilde{r}^{(l)}_{\bar{\mathcal{A}} n_{1}}(\lambda+\lambda_{n_{1}}^{(1)}-1)
\tilde{K}^{-(l,m)}_{\bar{\mathcal{A}}} (\lambda) \tilde{r}^{(l)}_{\bar{\mathcal{A}} n_{1}}(\lambda-\lambda_{n_{1}}^{(1)})
\dots \tilde{r}^{(l)}_{\bar{\mathcal{A}} 1}(\lambda-\lambda_{1}^{(1)}) \right] \nonumber \\
\end{eqnarray}
where $\tilde{r}^{(l)}_{\bar{\mathcal{A}} j}(\lambda) = \left( \mathcal{G}^{+(l)}_{\bar{\mathcal{A}}} \right)^{-1}
r_{\bar{\mathcal{A}} j}(\lambda)  \mathcal{G}^{+(l)}_{\bar{\mathcal{A}}}$ and
$\tilde{K}^{-(l,m)}_{\bar{\mathcal{A}}} (\lambda) =
\left(\mathcal{G}^{+(l)}_{\bar{\mathcal{A}}} \right)^{-1} \bar{K}^{-(m)}_{\bar{\mathcal{A}}} (\lambda)
\mathcal{G}^{+(l)}_{\bar{\mathcal{A}}}$.
Fortunetely in this case the gauge transformation on the operator
$r_{\bar{\mathcal{A}} j}(\lambda)$ can be reversed through the
transformation $r_{\bar{\mathcal{A}} j}(\lambda) = \left(\mathcal{G}^{+(l)}_{j} \right)^{-1}
\tilde{r}^{(l)}_{\bar{\mathcal{A}} j}(\lambda) \mathcal{G}^{+(l)}_{j}$.
Then the transfer matrix $\tilde{T}^{(l,m)}(\lambda,\{ \lambda^{(1)}_{j} \})$ defined as
\begin{eqnarray}
\label{tt}
\tilde{T}^{(l,m)}(\lambda,\{ \lambda^{(1)}_{j} \}) = \prod_{j=1}^{n_{1}} \left( \mathcal{G}^{+(l)}_{j} \right)^{-1} \;
\bar{T}^{(l,m)}(\lambda,\{ \lambda^{(1)}_{j} \})
\prod_{j=1}^{n_{1}} \mathcal{G}^{+(l)}_{j} 
\end{eqnarray}
is given by
\begin{eqnarray}
\label{gtnest}
&&\tilde{T}^{(l,m)}(\lambda,\{ \lambda^{(1)}_{j} \}) = \nonumber \\
&& \mbox{Tr}_{\bar{\mathcal{A}}}
\left[ D^{+(l)}_{\bar{\mathcal{A}}} (\lambda) r_{\bar{\mathcal{A}} 1}(\lambda+\lambda_{1}^{(1)}-1)
\dots r_{\bar{\mathcal{A}} n_{1}}(\lambda+\lambda_{n_{1}}^{(1)}-1)
\tilde{K}^{-(l,m)}_{\bar{\mathcal{A}}} (\lambda) r_{\bar{\mathcal{A}} n_{1}}(\lambda-\lambda_{n_{1}}^{(1)})
\dots r_{\bar{\mathcal{A}} 1}(\lambda-\lambda_{1}^{(1)}) \right]. \nonumber \\
\end{eqnarray}
Clearly, the operators $\tilde{T}^{(l,m)}(\lambda,\{ \lambda^{(1)}_{j} \})$
and $\bar{T}^{(l,m)}(\lambda,\{ \lambda^{(1)}_{j} \})$ share the same eigenvalues and
the eigenvectors $\ket{\tilde{\mathcal{F}}}$ of $\tilde{T}^{(l,m)}(\lambda,\{ \lambda^{(1)}_{j} \})$ are
related to $\ket{\mathcal{F}}$ by
\begin{equation}
\ket{\tilde{\mathcal{F}}} =\prod_{j=1}^{n_{1}} \left( \mathcal{G}^{+(l)}_{j} \right)^{-1} 
\ket{\mathcal{F}}.
\end{equation}
A careful examination of the matrix $\tilde{K}^{-(1,1)} (\lambda)$ reveals that it consists
of a non-diagonal matrix which still makes difficult to find an appropriate reference state to
perform a Bethe ansatz analysis. However, the requirement that $\tilde{K}^{-(1,1)} (\lambda)$
is upper or lower triangular allows the standard $sl(2)$ highest weight states to be used
as pseudovacuum state in the algebraic Bethe ansatz framework.
By requiring that $\tilde{K}^{-(1,1)} (\lambda)$ is upper triangular we find two possible classes
of restrictions for the boundary parameters, namely
\begin{eqnarray}
\label{c11a}
\mbox{C(1,1,a)}: \;\;\;\;\;\; h^{-}_{2} h^{+}_{4} &=& h^{-}_{4} h^{+}_{2} \\
\label{c11b}
\mbox{C(1,1,b)}: \;\;\;\;\;\; h^{-}_{1} h^{+}_{2} &=& h^{-}_{3} h^{+}_{4}.
\end{eqnarray}
On the other hand the matrix $\tilde{K}^{-(1,2)} (\lambda)$ is already diagonal and the diagonalization
of the double-row operator $\tilde{T}^{(1,2)}(\lambda,\{ \lambda^{(1)}_{j} \})$ can be performed
without any restriction for the boundary parameters. 

For the case $\bar{T}^{(2,1)}(\lambda,\{ \lambda^{(1)}_{j} \})$ we shall adopt a different strategy. In that
case we insert terms $\mathcal{G}^{-(1)}_{\bar{\mathcal{A}}} \left( \mathcal{G}^{-(1)}_{\bar{\mathcal{A}}} \right)^{-1}$
in between all the elements of the respective double-row operator (\ref{tnested}).
In this way the transfer matrix $\bar{T}^{(2,1)}(\lambda,\{ \lambda^{(1)}_{j} \})$ can be rewritten as
\begin{eqnarray}
\label{Ntnest1}
&&\bar{T}^{(2,1)}(\lambda,\{ \lambda^{(1)}_{j} \}) = \nonumber \\
&& \mbox{Tr}_{\bar{\mathcal{A}}}
\left[ \check{K}^{+(2,1)}_{\bar{\mathcal{A}}} (\lambda) \bar{r}^{(1)}_{\bar{\mathcal{A}} 1}(\lambda+\lambda_{1}^{(1)}-1)
\dots \bar{r}^{(1)}_{\bar{\mathcal{A}} n_{1}}(\lambda+\lambda_{n_{1}}^{(1)}-1)
D^{-(1)}_{\bar{\mathcal{A}}} (\lambda) \bar{r}^{(1)}_{\bar{\mathcal{A}} n_{1}}(\lambda-\lambda_{n_{1}}^{(1)})
\dots \bar{r}^{(1)}_{\bar{\mathcal{A}} 1}(\lambda-\lambda_{1}^{(1)}) \right] \nonumber \\
\end{eqnarray}
where $\bar{r}^{(1)}_{\bar{\mathcal{A}} j}(\lambda) = \left( \mathcal{G}^{-(1)}_{\bar{\mathcal{A}}} \right)^{-1}
r_{\bar{\mathcal{A}} j}(\lambda) \mathcal{G}^{-(1)}_{\bar{\mathcal{A}}}$ and
$\check{K}^{+(2,1)}_{\bar{\mathcal{A}}} (\lambda) =
\left( \mathcal{G}^{-(1)}_{\bar{\mathcal{A}}} \right)^{-1} \bar{K}^{+(2)}_{\bar{\mathcal{A}}} (\lambda)
\mathcal{G}^{-(1)}_{\bar{\mathcal{A}}}$.

In the same way as before (\ref{tt}) we can define a new transfer matrix
\begin{eqnarray}
\label{ctt}
\check{T}^{(2,1)}(\lambda,\{ \lambda^{(1)}_{j} \}) = \prod_{j=1}^{n_{1}} \left( \mathcal{G}^{-(1)}_{j} \right)^{-1} \;
\bar{T}^{(2,1)}(\lambda,\{ \lambda^{(1)}_{j} \})
\prod_{j=1}^{n_{1}} \mathcal{G}^{-(1)}_{j} 
\end{eqnarray}
sharing the same eigenvalues with $\bar{T}^{(2,1)}(\lambda,\{ \lambda^{(1)}_{j} \})$. The new double-row
operator $\check{T}^{(2,1)}(\lambda,\{ \lambda^{(1)}_{j} \})$ is given by
\begin{eqnarray}
\label{cgtnest}
&&\check{T}^{(2,1)}(\lambda,\{ \lambda^{(1)}_{j} \}) = \nonumber \\
&& \mbox{Tr}_{\bar{\mathcal{A}}}
\left[ \check{K}^{+(2,1)}_{\bar{\mathcal{A}}} (\lambda) r_{\bar{\mathcal{A}} 1}(\lambda+\lambda_{1}^{(1)}-1)
\dots r_{\bar{\mathcal{A}} n_{1}}(\lambda+\lambda_{n_{1}}^{(1)}-1)
D^{-(1)}_{\bar{\mathcal{A}}} (\lambda) r_{\bar{\mathcal{A}} n_{1}}(\lambda-\lambda_{n_{1}}^{(1)})
\dots r_{\bar{\mathcal{A}} 1}(\lambda-\lambda_{1}^{(1)}) \right] \nonumber \\
\end{eqnarray}
due to the relation
\begin{eqnarray}
\label{revgau1}
r_{\bar{\mathcal{A}} j}(\lambda) = \left( \mathcal{G}^{-(1)}_{j} \right)^{-1}
\bar{r}^{(1)}_{\bar{\mathcal{A}} j}(\lambda) \mathcal{G}^{-(1)}_{j} .
\end{eqnarray}
For instance, the matrix $\check{K}^{+(2,1)} (\lambda)$ is also diagonal and the eigenvalues
$\bar{\Lambda}^{(2,1)}(\lambda,\{ \lambda^{(1)}_{j} \})$ can be obtained for general values
of the boundary parameters. The remaining case $\bar{T}^{(2,2)}(\lambda,\{ \lambda^{(1)}_{j} \})$ 
contains only diagonal $K$-matrices and its diagonalization is straightforward through the algebraic
Bethe ansatz.
Considering that the diagonalization
of double-row transfer matrices with diagonal $K$-matrices has been well explained in the literature, see
for instance \cite{GONZA,PPP}, we restrict ourselves in presenting only the final solution for
$\Lambda^{(l,m)}(\lambda)$.
\begin{eqnarray}
\label{eigf}
&&\Lambda^{(l,m)}(\lambda) = Q^{(l,m)}_{1}(\lambda) a_{1}^{2L}(\lambda)
\prod_{i=1}^{n_{1}} \frac{(\lambda - \lambda^{(1)}_{i} + \frac{1}{2})}{(\lambda - \lambda^{(1)}_{i} -\frac{1}{2})}
\frac{(\lambda + \lambda^{(1)}_{i} + \frac{1}{2})}{(\lambda + \lambda^{(1)}_{i} -\frac{1}{2})} \nonumber \\
&&+Q^{(l,m)}_{2}(\lambda)  b^{2L}(\lambda)
\prod_{i=1}^{n_{1}} \frac{(\lambda - \lambda^{(1)}_{i} + \frac{1}{2})}{(\lambda - \lambda^{(1)}_{i} -\frac{1}{2})}
\frac{(\lambda + \lambda^{(1)}_{i} + \frac{1}{2})}{(\lambda + \lambda^{(1)}_{i} -\frac{1}{2})}
\prod_{i=1}^{n_{2}} \frac{(\lambda - \lambda^{(2)}_{i} - 1)}{(\lambda - \lambda^{(2)}_{i})}
\frac{(\lambda + \lambda^{(2)}_{i} - 1)}{(\lambda + \lambda^{(2)}_{i})} \nonumber \\
&&+ Q^{(l,m)}_{3}(\lambda) b^{2L}(\lambda)
\prod_{i=1}^{n_{2}} \frac{(\lambda - \lambda^{(2)}_{i} + 1)}{(\lambda - \lambda^{(2)}_{i})}
\frac{(\lambda + \lambda^{(2)}_{i} + 1)}{(\lambda + \lambda^{(2)}_{i})},
\end{eqnarray}
where the functions $Q^{(l,m)}_{i}(\lambda)$ are given by
\begin{eqnarray}
\label{funq11}
Q^{(1,1,a)}_{1}(\lambda)&=& -\frac{(\lambda+\frac{1}{2})}{(\lambda-\frac{1}{2})}
\frac{(\lambda + h_{0}^{+} - \frac{1}{2})}{(\lambda + h_{0}^{+} + \frac{1}{2})} 
\;\;\;\;\;
Q^{(1,1,b)}_{1}(\lambda) = -\frac{(\lambda+\frac{1}{2})}{(\lambda-\frac{1}{2})}
\frac{(\lambda + h_{0}^{+} - \frac{1}{2})}{(\lambda + h_{0}^{+} + \frac{1}{2})}
\\
Q^{(1,1,a)}_{2}(\lambda)&=&\frac{(\lambda+\frac{1}{2})}{(\lambda-\frac{1}{2})}
\frac{(\lambda + h_{0}^{+} - \frac{1}{2})}{(\lambda + h_{0}^{+} + \frac{1}{2})}
\;\;\;\;\;\;\;\;
 Q^{(1,1,b)}_{2}(\lambda)=\frac{(\lambda+\frac{1}{2})}{(\lambda-\frac{1}{2})}
\frac{(\lambda + h_{0}^{+} - \frac{1}{2})}{(\lambda + h_{0}^{+} + \frac{1}{2})}
\frac{(h_{0}^{-} +\lambda - 1)}{(h_{0}^{-} - \lambda )} \;\;\;\;\;\;\;\;
\\
Q^{(1,1,a)}_{3}(\lambda)&=&\frac{(\lambda + h_{0}^{-})}{(\lambda - h_{0}^{-})}
\frac{(\lambda - h_{0}^{+} + \frac{1}{2})}{(\lambda + h_{0}^{+} + \frac{1}{2})} 
\;\;\;\;\;\;
Q^{(1,1,b)}_{3}(\lambda)=\frac{(\lambda - h_{0}^{-} +1)}{(\lambda - h_{0}^{-})}
\frac{(h_{0}^{+} -\lambda - \frac{1}{2})}{(h_{0}^{+} +\lambda  + \frac{1}{2})}
\end{eqnarray}
\begin{eqnarray}
\label{funq12}
Q^{(1,2)}_{1}(\lambda)&=& \frac{(\lambda+\frac{1}{2})}{(\lambda-\frac{1}{2})}
\frac{(\lambda + h_{0}^{+} - \frac{1}{2})}{(\lambda + h_{0}^{+} + \frac{1}{2})}
\frac{(\lambda + h_{0}^{-})}{(\lambda - h_{0}^{-})}
\;\;\;\;\;\;\;\;\;\;\;\;\;
Q^{(2,1)}_{1}(\lambda) = \frac{(\lambda+\frac{1}{2})}{(\lambda-\frac{1}{2})}
\frac{(\lambda - h_{0}^{+} - \frac{3}{2})}{(\lambda + h_{0}^{+} + \frac{1}{2})} \;\;\;\;\;\;\;\;\;\;
\\
Q^{(1,2)}_{2}(\lambda)&=& \frac{(\lambda+\frac{1}{2})}{(\lambda-\frac{1}{2})}
\frac{(\lambda + h_{0}^{+} - \frac{1}{2})}{(\lambda + h_{0}^{+} + \frac{1}{2})}
\frac{(\lambda - h_{0}^{-} - 1)}{(\lambda - h_{0}^{-})}
\;\;\;\;\;\;\;
Q^{(2,1)}_{2}(\lambda) = \frac{(\lambda+\frac{1}{2})}{(\lambda-\frac{1}{2})} \\
Q^{(1,2)}_{3}(\lambda)&=& \frac{(\lambda - h_{0}^{-} - 1)}{(\lambda - h_{0}^{-})}
\frac{(h_{0}^{+} -\lambda - \frac{1}{2})}{(h_{0}^{+} +\lambda + \frac{1}{2})}
\;\;\;\;\;\;\;\;\;\;\;\;\;\;\;\;\;\;\;
Q^{(2,1)}_{3}(\lambda) = \frac{(h_{0}^{-} +\lambda )}{(h_{0}^{-} - \lambda)}
\end{eqnarray}
\begin{eqnarray}
\label{funq22}
Q^{(2,2)}_{1}(\lambda)&=& \frac{(\lambda+\frac{1}{2})}{(\lambda-\frac{1}{2})}
\frac{(\lambda - h_{0}^{+} - \frac{3}{2})}{(\lambda + h_{0}^{+} + \frac{1}{2})}
\frac{(h_{0}^{-} +\lambda)}{(h_{0}^{-} -\lambda)} \\
Q^{(2,2)}_{2}(\lambda)&=&  \frac{(\lambda+\frac{1}{2})}{(\lambda-\frac{1}{2})}
\frac{(\lambda - h_{0}^{-} - 1)}{(\lambda - h_{0}^{-} )} \\
Q^{(2,2)}_{3}(\lambda)&=&  \frac{(\lambda - h_{0}^{-} -1)}{(\lambda - h_{0}^{-})}
\label{qf}
\end{eqnarray}
In order to capture the two classes of restrictions (\ref{c11a},\ref{c11b}),
in which we are able to present the eigenvalue $\Lambda^{(1,1)}(\lambda)$, we have used 
$Q^{(1,1,a)}_{i}(\lambda)$ and $Q^{(1,1,b)}_{i}(\lambda)$ to denote the 
functions $Q^{(1,1)}_{i}(\lambda)$ obtained under the constraints C(1,1,a) and C(1,1,b)
respectively.
The set of rapidities $\{ \lambda^{(2)}_{1}, \dots , \lambda^{(2)}_{n_{2}} \}$, introduced in the
diagonalization of $\bar{T}^{(l,m)}(\lambda,\{ \lambda^{(1)}_{j} \})$, together with the set
$\{ \lambda^{(1)}_{1}, \dots , \lambda^{(1)}_{n_{1}} \}$ are required to satisfy the following 
Bethe ansatz equations,
\begin{eqnarray}
\label{banf}
\left( \frac{\lambda^{(1)}_{j} - \frac{1}{2}}{\lambda^{(1)}_{j} + \frac{1}{2}} \right)^{2L} 
\Theta^{(l,m)}_{1}(\lambda^{(1)}_{j}) &=&
\prod_{i=1}^{n_{2}} \frac{(\lambda^{(1)}_{j} - \lambda^{(2)}_{i} - \frac{1}{2})}{(\lambda^{(1)}_{j} - \lambda^{(2)}_{i} + \frac{1}{2})}
\frac{(\lambda^{(1)}_{j} + \lambda^{(2)}_{i} - \frac{1}{2})}{(\lambda^{(1)}_{j} + \lambda^{(2)}_{i} + \frac{1}{2})} \nonumber \\
\prod_{i=1}^{n_{1}}
\frac{(\lambda^{(2)}_{j} - \lambda^{(1)}_{i} + \frac{1}{2})}{(\lambda^{(2)}_{j} - \lambda^{(1)}_{i} - \frac{1}{2})}
\frac{(\lambda^{(2)}_{j} + \lambda^{(1)}_{i} + \frac{1}{2})}{(\lambda^{(2)}_{j} + \lambda^{(1)}_{i} - \frac{1}{2})}
\Theta^{(l,m)}_{2}(\lambda^{(2)}_{j}) &=& \prod_{i\neq j}^{n_{2}} \frac{(\lambda^{(1)}_{j} - \lambda^{(1)}_{i} + 1)}{(\lambda^{(1)}_{j} - \lambda^{(1)}_{i} -1)}
\frac{(\lambda^{(1)}_{j} + \lambda^{(1)}_{i} + 1)}{(\lambda^{(1)}_{j} + \lambda^{(1)}_{i} -1)}. \nonumber \\
\end{eqnarray}
The functions $\Theta^{(l,m)}_{i}(\lambda)$ are given by
\begin{eqnarray}
\label{baf1}
\Theta^{(1,1,a)}_{1}(\lambda)&=& 1 
\;\;\;\;\;\;\;\;\;\;\;\;\;\;\;\;\;\;\;\;\;\;\;\;\;\;\;\;\;\;\;\;\;\;\;\;\;\;
\Theta^{(1,1,a)}_{2}(\lambda)= \frac{(\lambda + h_{0}^{+} - \frac{1}{2})}{(\lambda - h_{0}^{+} + \frac{1}{2})}
\frac{(\lambda - h_{0}^{-})}{(\lambda + h_{0}^{-})} \\
\Theta^{(1,1,b)}_{1}(\lambda) &=& \frac{(h_{0}^{-} -\lambda - \frac{1}{2})}{(h_{0}^{-} +\lambda - \frac{1}{2})}
\;\;\;\;\;\;\;\;\;\;\;\;\;\;\;\;\;\;\;\;
\Theta^{(1,1,b)}_{2}(\lambda)=\frac{(\lambda + h_{0}^{+} - \frac{1}{2})}{(\lambda - h_{0}^{+} + \frac{1}{2})}
\frac{(\lambda + h_{0}^{-} -1)}{(\lambda - h_{0}^{-} + 1)} \;\;\;\;\;\;\;
\end{eqnarray}
\begin{eqnarray}
\label{baf2}
\Theta^{(1,2)}_{1}(\lambda)&=& \frac{(h_{0}^{-} +\lambda + \frac{1}{2})}{(h_{0}^{-} -\lambda - \frac{1}{2})}
\;\;\;\;\;\;\;\;\;\;\;\;\;\;\;\;\;\;\;\;\;\;\;\;\;\;\;\;\;\;\;\;\;\;\;\;\;\;\;
\Theta^{(1,2)}_{2}(\lambda) = \frac{(h_{0}^{+} +\lambda - \frac{1}{2})}{(h_{0}^{+} - \lambda - \frac{1}{2})}
\;\;\;\;\;\;\;\;\;\;\;\;\;\;\;\; \\
\Theta^{(2,1)}_{1}(\lambda)&=& \frac{(h_{0}^{+} -\lambda + 1)}{(h_{0}^{+} +\lambda + 1)}
\;\;\;\;\;\;\;\;\;\;\;\;\;\;\;\;\;\;\;\;\;\;\;\;\;\;\;\;\;\;\;\;\;\;\;\;\;\;\;\;
\Theta^{(2,1)}_{2}(\lambda) = \frac{(h_{0}^{-} - \lambda)}{(h_{0}^{-} + \lambda)} \\
\Theta^{(2,2)}_{1}(\lambda)&=& \frac{(\lambda - h_{0}^{+} - 1)}{(\lambda + h_{0}^{+} + 1)}
\frac{(\lambda + h_{0}^{-} + \frac{1}{2})}{(\lambda - h_{0}^{-} - \frac{1}{2})}
\;\;\;\;\;\;\;\;\;\;\;\;\;\;\;\;\;\;\;\;
\Theta^{(2,2)}_{2}(\lambda) = 1
\label{ttf}
\end{eqnarray}
where $\Theta^{(1,1,a)}_{i}(\lambda)$ and $\Theta^{(1,1,b)}_{i}(\lambda)$ denote the functions
$\Theta^{(1,1)}_{i}(\lambda)$ under the constraints C(1,1,a) and C(1,1,b) respectively.

Considering our results so far, the eigenvalues $E^{(l,m)}$ of the
hamiltonian $\mathcal{H}^{(l,m)}$ are given by
\begin{equation}
\label{elm}
E^{(l,m)} = - \sum_{j=1}^{n_{1}} \frac{1}{(\lambda^{(1)}_{j})^{2} - \frac{1}{4}},
\end{equation}
where the rapidities $\lambda^{(1)}_{j}$ satisfy the Bethe ansatz equations (\ref{banf})
with the corresponding index $(l,m)$.

\section{Concluding Remarks}

In this work we have constructed supersymmetric t-J models with integrable open boundaries
through the Quantum Inverse Scattering Method. Four different kinds of open boundaries are
obtained: one having only diagonal elements, two with one diagonal boundary and the other
non-diagonal, and one with two non-diagonal boundaries. The exact solution of the corresponding
models is obtained by means of the algebraic Bethe ansatz.
We also showed that the covering transfer matrix of the supersymmetric
t-J model built from a general non-diagonal solution of the reflection equation possess a trivial reference
state. Furthermore, we also presented the one-particle eigenvalue of the corresponding
transfer matrix without any restriction for the boundary parameters.

One interesting possibility that deserves attention is the generalization of the above results for the
$q$-deformed t-J model based on the $U_{q}[sl(2|1)]$ symmetry. Here we remark that progresses have been
reported for the XXZ model with non-diagonal open boundaries \cite{NEPO}-\cite{GIER} and the extension of these
results for the $q$-deformed t-J model deserves to be investigated.

\section{Acknowledgements}
W. Galleas thanks M.J. Martins for useful discussions and Fapesp for finantial support.

\newpage
\section*{\bf Appendix A: Auxiliary functions}
\setcounter{equation}{0}
\renewcommand{\theequation}{A.\arabic{equation}}
In this appendix we present the functions $X^{(l,m)}_{i}$ required in
section 3 for the diagonalization of the $n_{1}=1$ nested problem.
\begin{eqnarray}
\label{coef1}
X^{(l,m)}_{2} &=& \left(1 - 2\lambda \right)^2  \\ \nonumber \\
X^{(l,m)}_{1} &=& \left(1 - 2\lambda \right) \left\{
\left( 4 Y^{(l,m)}_{3} \lambda + 2Y^{(l,m)}_{2} \right)
\left[ \lambda^{2} + (1 - \lambda^{(1)}_{1}) \lambda^{(1)}_{1} \right]
+ 2 Y^{(l,m)}_{1} \lambda + Y^{(l,m)}_{0} \right\}  \\  \nonumber \\
X^{(l,m)}_{0} &=&
\lambda \left( 4 (Y^{(l,m)}_{3})^{2} \lambda + 4 Y^{(l,m)}_{2} Y^{(l,m)}_{3}  \right)
\left[ \lambda^{4} + 2 \lambda^{(1)}_{1} (1-\lambda^{(1)}_{1}) \lambda^2
+ (\lambda^{(1)}_{1})^{3} (\lambda^{(1)}_{1} -2) \right] \nonumber \\
&+& (\lambda^{(1)}_{1})^{2} \left( -2 Z^{(l,m)}_{10} \lambda^{2} + 2 Z^{(l,m)}_{9} \lambda + Z^{(l,m)}_{8} \right)
+ \lambda^{(1)}_{1} \left( 2 Z^{(l,m)}_{7} \lambda^{2} + 2 Z^{(l,m)}_{6} \lambda + Z^{(l,m)}_{5} \right) \nonumber \\
&+& Z^{(l,m)}_{4} \lambda^{4} -2 Z^{(l,m)}_{3} \lambda^{3} + Z^{(l,m)}_{2} \lambda^{2}
+ Z^{(l,m)}_{1} \lambda + Z^{(l,m)}_{0}
\end{eqnarray}
The functions $Y^{(l,m)}_{i}$ and $Z^{(l,m)}_{i}$ contain the dependence on the elements of the
$K$-matrices and they are explicitly given by
\begin{eqnarray}
Y^{(l,m)}_{3} &=&  k^{-(m)}_{22}(\lambda)k^{+(l)}_{22}(\lambda) + k^{-(m)}_{32}(\lambda)k^{+(l)}_{23}(\lambda) + k^{-(m)}_{23}(\lambda)k^{+(l)}_{32}(\lambda) +
k^{-(m)}_{33}(\lambda)k^{+(l)}_{33}(\lambda) \nonumber \\
\\
Y^{(l,m)}_{2} &=& k^{-(m)}_{11}(\lambda)k^{+(l)}_{22}(\lambda) - k^{-(m)}_{22}(\lambda)k^{+(l)}_{22}(\lambda) - k^{-(m)}_{32}(\lambda)k^{+(l)}_{23}(\lambda) -
k^{-(m)}_{23}(\lambda)k^{+(l)}_{32}(\lambda) \nonumber \\
&+& k^{-(m)}_{11}(\lambda)k^{+(l)}_{33}(\lambda) - k^{-(m)}_{33}(\lambda)k^{+(l)}_{33}(\lambda) \\
\nonumber \\
Y^{(l,m)}_{1} &=& k^{-(m)}_{33}(\lambda)k^{+(l)}_{22}(\lambda) - k^{-(m)}_{32}(\lambda)k^{+(l)}_{23}(\lambda) - k^{-(m)}_{23}(\lambda)k^{+(l)}_{32}(\lambda) +
k^{-(m)}_{22}(\lambda)k^{+(l)}_{33}(\lambda) \nonumber \\
\\
Y^{(l,m)}_{0} &=& k^{-(m)}_{11}(\lambda)k^{+(l)}_{22}(\lambda) - k^{-(m)}_{33}(\lambda)k^{+(l)}_{22}(\lambda) + k^{-(m)}_{32}(\lambda)k^{+(l)}_{23}(\lambda) +
k^{-(m)}_{23}(\lambda)k^{+(l)}_{32}(\lambda) \nonumber \\
&+& k^{-(m)}_{11}(\lambda)k^{+(l)}_{33}(\lambda) - k^{-(m)}_{22}(\lambda)k^{+(l)}_{33}(\lambda)
\end{eqnarray}
\begin{eqnarray}
Z^{(l,m)}_{10} &=& (k^{-(m)}_{11}(\lambda))^{2}(k^{+(l)}_{22}(\lambda))^{2} - 2k^{-(m)}_{11}(\lambda)k^{-(m)}_{22}(\lambda)(k^{+(l)}_{22}(\lambda))^{2} -
(k^{-(m)}_{22}(\lambda))^{2}(k^{+(l)}_{22}(\lambda))^{2} \nonumber \\
&-& 2k^{-(m)}_{23}(\lambda)k^{-(m)}_{32}(\lambda)(k^{+(l)}_{22}(\lambda))^{2} +
2k^{-(m)}_{22}(\lambda)k^{-(m)}_{33}(\lambda)(k^{+(l)}_{22}(\lambda))^{2}
\nonumber \\
&-& 2k^{-(m)}_{22}(\lambda)k^{-(m)}_{32}(\lambda)k^{+(l)}_{22}(\lambda)k^{+(l)}_{23}(\lambda) -
(k^{-(m)}_{32}(\lambda))^{2}(k^{+(l)}_{23}(\lambda))^{2}
\nonumber \\
&-& 2k^{-(m)}_{11}(\lambda)k^{-(m)}_{23}(\lambda)k^{+(l)}_{22}(\lambda)k^{+(l)}_{32}(\lambda)
-2k^{-(m)}_{22}(\lambda)k^{-(m)}_{23}(\lambda)k^{+(l)}_{22}(\lambda)k^{+(l)}_{32}(\lambda)
\nonumber \\
&-& 2(k^{-(m)}_{22}(\lambda))^{2}k^{+(l)}_{23}(\lambda)
k^{+(l)}_{32}(\lambda) - 10k^{-(m)}_{23}(\lambda)k^{-(m)}_{32}(\lambda)k^{+(l)}_{23}(\lambda)k^{+(l)}_{32}(\lambda)
\nonumber \\
&+&4k^{-(m)}_{22}(\lambda)k^{-(m)}_{33}(\lambda)k^{+(l)}_{23}(\lambda)k^{+(l)}_{32}(\lambda)
- 2k^{-(m)}_{11}(\lambda)k^{-(m)}_{32}(\lambda)k^{+(l)}_{22}(\lambda)k^{+(l)}_{23}(\lambda)
\nonumber \\
&-& 2(k^{-(m)}_{33}(\lambda))^{2}k^{+(l)}_{23}(\lambda)
k^{+(l)}_{32}(\lambda) - (k^{-(m)}_{23}(\lambda))^{2}(k^{+(l)}_{32}(\lambda))^{2}
\nonumber \\
&+& 2(k^{-(m)}_{11}(\lambda))^{2}k^{+(l)}_{22}(\lambda)
k^{+(l)}_{33}(\lambda) - 2k^{-(m)}_{11}(\lambda)k^{-(m)}_{22}(\lambda)k^{+(l)}_{22}(\lambda)k^{+(l)}_{33}(\lambda)
\nonumber \\
&+& 2(k^{-(m)}_{22}(\lambda))^{2}k^{+(l)}_{22}(\lambda)k^{+(l)}_{33}(\lambda)
+ 4k^{-(m)}_{23}(\lambda)k^{-(m)}_{32}(\lambda)k^{+(l)}_{22}(\lambda)
k^{+(l)}_{33}(\lambda)
\nonumber \\
&-& 2k^{-(m)}_{11}(\lambda)k^{-(m)}_{33}(\lambda)k^{+(l)}_{22}(\lambda)k^{+(l)}_{33}(\lambda) -
2k^{-(m)}_{22}(\lambda)k^{-(m)}_{33}(\lambda)k^{+(l)}_{22}(\lambda)k^{+(l)}_{33}(\lambda)
\nonumber \\
&+& 2(k^{-(m)}_{33}(\lambda))^{2}k^{+(l)}_{22}(\lambda)
k^{+(l)}_{33}(\lambda) - 2k^{-(m)}_{11}(\lambda)k^{-(m)}_{32}(\lambda)k^{+(l)}_{23}(\lambda)k^{+(l)}_{33}(\lambda)
\nonumber \\
&-&2k^{-(m)}_{32}(\lambda)k^{-(m)}_{33}(\lambda)k^{+(l)}_{23}(\lambda)k^{+(l)}_{33}(\lambda)
- 2k^{-(m)}_{11}(\lambda)k^{-(m)}_{23}(\lambda)k^{+(l)}_{32}(\lambda)k^{+(l)}_{33}(\lambda)
\nonumber \\
&-& 2k^{-(m)}_{23}(\lambda)k^{-(m)}_{33}(\lambda)k^{+(l)}_{32}(\lambda)k^{+(l)}_{33}(\lambda) +
(k^{-(m)}_{11}(\lambda))^{2}(k^{+(l)}_{33}(\lambda))^{2}
\nonumber \\
&-& 2k^{-(m)}_{23}(\lambda)k^{-(m)}_{32}(\lambda)(k^{+(l)}_{33}(\lambda))^{2} -
2k^{-(m)}_{11}(\lambda)k^{-(m)}_{33}(\lambda)(k^{+(l)}_{33}(\lambda))^{2}
\nonumber \\
&+& 2k^{-(m)}_{22}(\lambda)k^{-(m)}_{33}(\lambda)(k^{+(l)}_{33}(\lambda))^{2} -
(k^{-(m)}_{33}(\lambda))^{2}(k^{+(l)}_{33}(\lambda))^{2}
\end{eqnarray}
\begin{eqnarray}
Z^{(l,m)}_{9} &=& k^{-(m)}_{11}(\lambda)k^{-(m)}_{22}(\lambda)(k^{+(l)}_{22}(\lambda))^{2} - 2(k^{-(m)}_{22}(\lambda))^{2}(k^{+(l)}_{22}(\lambda))^{2} -
2k^{-(m)}_{23}(\lambda)k^{-(m)}_{32}(\lambda)(k^{+(l)}_{22}(\lambda))^{2}
\nonumber \\
&-& k^{-(m)}_{11}(\lambda)k^{-(m)}_{33}(\lambda)(k^{+(l)}_{22}(\lambda))^{2} +
2k^{-(m)}_{22}(\lambda)k^{-(m)}_{33}(\lambda)(k^{+(l)}_{22}(\lambda))^{2}
\nonumber \\
&+& 2k^{-(m)}_{11}(\lambda)k^{-(m)}_{32}(\lambda)k^{+(l)}_{22}(\lambda)
k^{+(l)}_{23}(\lambda) - 4k^{-(m)}_{22}(\lambda)k^{-(m)}_{32}(\lambda)k^{+(l)}_{22}(\lambda)k^{+(l)}_{23}(\lambda)
\nonumber \\
&-&2(k^{-(m)}_{32}(\lambda))^{2}(k^{+(l)}_{23}(\lambda))^{2} + 2k^{-(m)}_{11}(\lambda)k^{-(m)}_{23}(\lambda)k^{+(l)}_{22}(\lambda)k^{+(l)}_{32}(\lambda)
\nonumber \\
&-&4k^{-(m)}_{22}(\lambda)k^{-(m)}_{23}(\lambda)k^{+(l)}_{22}(\lambda)k^{+(l)}_{32}(\lambda) - 2(k^{-(m)}_{22}(\lambda))^{2}k^{+(l)}_{23}(\lambda)
k^{+(l)}_{32}(\lambda)
\nonumber \\
&-& 12k^{-(m)}_{23}(\lambda)k^{-(m)}_{32}(\lambda)k^{+(l)}_{23}(\lambda)k^{+(l)}_{32}(\lambda) +
4k^{-(m)}_{22}(\lambda)k^{-(m)}_{33}(\lambda)k^{+(l)}_{23}(\lambda)k^{+(l)}_{32}(\lambda)
\nonumber \\
&-& 2(k^{-(m)}_{33}(\lambda))^{2}k^{+(l)}_{23}(\lambda)
k^{+(l)}_{32}(\lambda) - 2(k^{-(m)}_{23}(\lambda))^{2}(k^{+(l)}_{32}(\lambda))^{2}
+ 2(k^{-(m)}_{22}(\lambda))^{2}k^{+(l)}_{22}(\lambda)k^{+(l)}_{33}(\lambda)
\nonumber \\
&+& 4k^{-(m)}_{23}(\lambda)k^{-(m)}_{32}(\lambda)k^{+(l)}_{22}(\lambda)k^{+(l)}_{33}(\lambda) -
4k^{-(m)}_{22}(\lambda)k^{-(m)}_{33}(\lambda)k^{+(l)}_{22}(\lambda)k^{+(l)}_{33}(\lambda)
\nonumber \\
&+& 2(k^{-(m)}_{33}(\lambda))^{2}k^{+(l)}_{22}(\lambda)
k^{+(l)}_{33}(\lambda) + 2k^{-(m)}_{11}(\lambda)k^{-(m)}_{32}(\lambda)k^{+(l)}_{23}(\lambda)k^{+(l)}_{33}(\lambda)
\nonumber \\
&-& 4k^{-(m)}_{32}(\lambda)k^{-(m)}_{33}(\lambda)k^{+(l)}_{23}(\lambda)k^{+(l)}_{33}(\lambda) + 2k^{-(m)}_{11}(\lambda)k^{-(m)}_{23}(\lambda)
k^{+(l)}_{32}(\lambda)k^{+(l)}_{33}(\lambda)
\nonumber \\
&-& 4k^{-(m)}_{23}(\lambda)k^{-(m)}_{33}(\lambda)k^{+(l)}_{32}(\lambda)k^{+(l)}_{33}(\lambda) -
k^{-(m)}_{11}(\lambda)k^{-(m)}_{22}(\lambda)(k^{+(l)}_{33}(\lambda))^{2}
\nonumber \\
&-& 2k^{-(m)}_{23}(\lambda)k^{-(m)}_{32}(\lambda)(k^{+(l)}_{33}(\lambda))^{2} +
k^{-(m)}_{11}(\lambda)k^{-(m)}_{33}(\lambda)(k^{+(l)}_{33}(\lambda))^{2}
\nonumber \\
&+& 2k^{-(m)}_{22}(\lambda)k^{-(m)}_{33}(\lambda)(k^{+(l)}_{33}(\lambda))^{2} -
2(k^{-(m)}_{33}(\lambda))^{2}(k^{+(l)}_{33}(\lambda))^{2}
\end{eqnarray}
\begin{eqnarray}
Z^{(l,m)}_{8} &=& -k^{-(m)}_{11}(\lambda)k^{-(m)}_{22}(\lambda)(k^{+(l)}_{22}(\lambda))^{2} + (k^{-(m)}_{22}(\lambda))^{2}(k^{+(l)}_{22}(\lambda))^{2} +
k^{-(m)}_{23}(\lambda)k^{-(m)}_{32}(\lambda)(k^{+(l)}_{22}(\lambda))^{2}
\nonumber \\
&+& k^{-(m)}_{11}(\lambda)k^{-(m)}_{33}(\lambda)(k^{+(l)}_{22}(\lambda))^{2} -
k^{-(m)}_{22}(\lambda)k^{-(m)}_{33}(\lambda)(k^{+(l)}_{22}(\lambda))^{2}
\nonumber \\
&-& 2k^{-(m)}_{11}(\lambda)k^{-(m)}_{32}(\lambda)k^{+(l)}_{22}(\lambda)
k^{+(l)}_{23}(\lambda) + 2k^{-(m)}_{22}(\lambda)k^{-(m)}_{32}(\lambda)k^{+(l)}_{22}(\lambda)k^{+(l)}_{23}(\lambda)
\nonumber \\
&+&(k^{-(m)}_{32}(\lambda))^{2}(k^{+(l)}_{23}(\lambda))^{2} - 2k^{-(m)}_{11}(\lambda)k^{-(m)}_{23}(\lambda)k^{+(l)}_{22}(\lambda)k^{+(l)}_{32}(\lambda)
\nonumber \\
&+& 2k^{-(m)}_{22}(\lambda)k^{-(m)}_{23}(\lambda)k^{+(l)}_{22}(\lambda)k^{+(l)}_{32}(\lambda) + (k^{-(m)}_{22}(\lambda))^{2}k^{+(l)}_{23}(\lambda)
k^{+(l)}_{32}(\lambda)
\nonumber \\
&+& 6k^{-(m)}_{23}(\lambda)k^{-(m)}_{32}(\lambda)k^{+(l)}_{23}(\lambda)k^{+(l)}_{32}(\lambda) -
2k^{-(m)}_{22}(\lambda)k^{-(m)}_{33}(\lambda)k^{+(l)}_{23}(\lambda)k^{+(l)}_{32}(\lambda)
\nonumber \\
&+& (k^{-(m)}_{33}(\lambda))^{2}k^{+(l)}_{23}(\lambda)
k^{+(l)}_{32}(\lambda) + (k^{-(m)}_{23}(\lambda))^{2}(k^{+(l)}_{32}(\lambda))^{2} - (k^{-(m)}_{22}(\lambda))^{2}k^{+(l)}_{22}(\lambda)k^{+(l)}_{33}(\lambda)
\nonumber \\
&-&2k^{-(m)}_{23}(\lambda)k^{-(m)}_{32}(\lambda)k^{+(l)}_{22}(\lambda)k^{+(l)}_{33}(\lambda) + 2k^{-(m)}_{22}(\lambda)k^{-(m)}_{33}(\lambda)
k^{+(l)}_{22}(\lambda)k^{+(l)}_{33}(\lambda)
\nonumber \\
&-& (k^{-(m)}_{33}(\lambda))^{2}k^{+(l)}_{22}(\lambda)k^{+(l)}_{33}(\lambda) -
2k^{-(m)}_{11}(\lambda)k^{-(m)}_{32}(\lambda)k^{+(l)}_{23}(\lambda)k^{+(l)}_{33}(\lambda)
\nonumber \\
&+& 2k^{-(m)}_{32}(\lambda)k^{-(m)}_{33}(\lambda)
k^{+(l)}_{23}(\lambda)k^{+(l)}_{33}(\lambda) - 2k^{-(m)}_{11}(\lambda)k^{-(m)}_{23}(\lambda)k^{+(l)}_{32}(\lambda)k^{+(l)}_{33}(\lambda)
\nonumber \\
&+& 2k^{-(m)}_{23}(\lambda)k^{-(m)}_{33}(\lambda)k^{+(l)}_{32}(\lambda)k^{+(l)}_{33}(\lambda) + k^{-(m)}_{11}(\lambda)k^{-(m)}_{22}(\lambda)
(k^{+(l)}_{33}(\lambda))^{2}
\nonumber \\
&+& k^{-(m)}_{23}(\lambda)k^{-(m)}_{32}(\lambda)(k^{+(l)}_{33}(\lambda))^{2} -
k^{-(m)}_{11}(\lambda)k^{-(m)}_{33}(\lambda)(k^{+(l)}_{33}(\lambda))^{2}
\nonumber \\
&-& k^{-(m)}_{22}(\lambda)k^{-(m)}_{33}(\lambda)(k^{+(l)}_{33}(\lambda))^{2} +
(k^{-(m)}_{33}(\lambda))^{2}(k^{+(l)}_{33}(\lambda))^{2}
\end{eqnarray}
\begin{eqnarray}
Z^{(l,m)}_{7} &=& (k^{-(m)}_{11}(\lambda))^{2}(k^{+(l)}_{22}(\lambda))^{2} - 2k^{-(m)}_{11}(\lambda)k^{-(m)}_{22}(\lambda)(k^{+(l)}_{22}(\lambda))^{2} +
(k^{-(m)}_{22}(\lambda))^{2}(k^{+(l)}_{22}(\lambda))^{2}
\nonumber \\
&-& 2k^{-(m)}_{23}(\lambda)k^{-(m)}_{32}(\lambda)(k^{+(l)}_{22}(\lambda))^{2} +
2k^{-(m)}_{22}(\lambda)k^{-(m)}_{33}(\lambda)(k^{+(l)}_{22}(\lambda))^{2}
\nonumber \\
&-& 2k^{-(m)}_{11}(\lambda)k^{-(m)}_{32}(\lambda)k^{+(l)}_{22}(\lambda)
k^{+(l)}_{23}(\lambda) + 2k^{-(m)}_{22}(\lambda)k^{-(m)}_{32}(\lambda)k^{+(l)}_{22}(\lambda)k^{+(l)}_{23}(\lambda)
\nonumber \\
&+& (k^{-(m)}_{32}(\lambda))^{2}(k^{+(l)}_{23}(\lambda))^{2} - 2k^{-(m)}_{11}(\lambda)k^{-(m)}_{23}(\lambda)k^{+(l)}_{22}(\lambda)k^{+(l)}_{32}(\lambda)
\nonumber \\
&+& 2k^{-(m)}_{22}(\lambda)k^{-(m)}_{23}(\lambda)k^{+(l)}_{22}(\lambda)k^{+(l)}_{32}(\lambda) - 2(k^{-(m)}_{22}(\lambda))^{2}k^{+(l)}_{23}(\lambda)
k^{+(l)}_{32}(\lambda)
\nonumber \\
&-& 6k^{-(m)}_{23}(\lambda)k^{-(m)}_{32}(\lambda)k^{+(l)}_{23}(\lambda)k^{+(l)}_{32}(\lambda) +
4k^{-(m)}_{22}(\lambda)k^{-(m)}_{33}(\lambda)k^{+(l)}_{23}(\lambda)k^{+(l)}_{32}(\lambda)
\nonumber \\
&-& 2(k^{-(m)}_{33}(\lambda))^{2}k^{+(l)}_{23}(\lambda)
k^{+(l)}_{32}(\lambda) + (k^{-(m)}_{23}(\lambda))^{2}(k^{+(l)}_{32}(\lambda))^{2} + 2(k^{-(m)}_{11}(\lambda))^{2}k^{+(l)}_{22}(\lambda)
k^{+(l)}_{33}(\lambda)
\nonumber \\
&-& 2k^{-(m)}_{11}(\lambda)k^{-(m)}_{22}(\lambda)k^{+(l)}_{22}(\lambda)k^{+(l)}_{33}(\lambda) +
2(k^{-(m)}_{22}(\lambda))^{2}k^{+(l)}_{22}(\lambda)k^{+(l)}_{33}(\lambda)
\nonumber \\
&+& 4k^{-(m)}_{23}(\lambda)k^{-(m)}_{32}(\lambda)k^{+(l)}_{22}(\lambda)
k^{+(l)}_{33}(\lambda) - 2k^{-(m)}_{11}(\lambda)k^{-(m)}_{33}(\lambda)k^{+(l)}_{22}(\lambda)k^{+(l)}_{33}(\lambda)
\nonumber \\
&+& 2k^{-(m)}_{22}(\lambda)k^{-(m)}_{33}(\lambda)k^{+(l)}_{22}(\lambda)k^{+(l)}_{33}(\lambda) + 2(k^{-(m)}_{33}(\lambda))^{2}k^{+(l)}_{22}(\lambda)
k^{+(l)}_{33}(\lambda)
\nonumber \\
&-& 2k^{-(m)}_{11}(\lambda)k^{-(m)}_{32}(\lambda)k^{+(l)}_{23}(\lambda)k^{+(l)}_{33}(\lambda) +
2k^{-(m)}_{32}(\lambda)k^{-(m)}_{33}(\lambda)k^{+(l)}_{23}(\lambda)k^{+(l)}_{33}(\lambda)
\nonumber \\
&-& 2k^{-(m)}_{11}(\lambda)k^{-(m)}_{23}(\lambda)
k^{+(l)}_{32}(\lambda)k^{+(l)}_{33}(\lambda) + 2k^{-(m)}_{23}(\lambda)k^{-(m)}_{33}(\lambda)k^{+(l)}_{32}(\lambda)k^{+(l)}_{33}(\lambda)
\nonumber \\
&+& (k^{-(m)}_{11}(\lambda))^{2}(k^{+(l)}_{33}(\lambda))^{2} - 2k^{-(m)}_{23}(\lambda)k^{-(m)}_{32}(\lambda)(k^{+(l)}_{33}(\lambda))^{2} -
2k^{-(m)}_{11}(\lambda)k^{-(m)}_{33}(\lambda)(k^{+(l)}_{33}(\lambda))^{2}
\nonumber \\
&+& 2k^{-(m)}_{22}(\lambda)k^{-(m)}_{33}(\lambda)(k^{+(l)}_{33}(\lambda))^{2} +
(k^{-(m)}_{33}(\lambda))^{2}(k^{+(l)}_{33}(\lambda))^{2}
\end{eqnarray}
\begin{eqnarray}
Z^{(l,m)}_{6} &=& k^{-(m)}_{11}(\lambda)k^{-(m)}_{22}(\lambda)(k^{+(l)}_{22}(\lambda))^{2} + 2k^{-(m)}_{23}(\lambda)k^{-(m)}_{32}(\lambda)(k^{+(l)}_{22}(\lambda))^{2}
\nonumber \\
&+& k^{-(m)}_{11}(\lambda)k^{-(m)}_{33}(\lambda)(k^{+(l)}_{22}(\lambda))^{2}
- 2k^{-(m)}_{22}(\lambda)k^{-(m)}_{33}(\lambda)(k^{+(l)}_{22}(\lambda))^{2}
\nonumber \\
&+& 2(k^{-(m)}_{22}(\lambda))^{2}k^{+(l)}_{23}(\lambda)k^{+(l)}_{32}(\lambda) + 8k^{-(m)}_{23}(\lambda)k^{-(m)}_{32}(\lambda)k^{+(l)}_{23}(\lambda)
k^{+(l)}_{32}(\lambda)
\nonumber \\
&-& 4k^{-(m)}_{22}(\lambda)k^{-(m)}_{33}(\lambda)k^{+(l)}_{23}(\lambda)k^{+(l)}_{32}(\lambda) +
2(k^{-(m)}_{33}(\lambda))^{2}k^{+(l)}_{23}(\lambda)k^{+(l)}_{32}(\lambda)
\nonumber \\
&+& 2k^{-(m)}_{11}(\lambda)k^{-(m)}_{22}(\lambda)k^{+(l)}_{22}(\lambda)
k^{+(l)}_{33}(\lambda) - 2(k^{-(m)}_{22}(\lambda))^{2}k^{+(l)}_{22}(\lambda)k^{+(l)}_{33}(\lambda)
\nonumber \\
&-& 4k^{-(m)}_{23}(\lambda)k^{-(m)}_{32}(\lambda)k^{+(l)}_{22}(\lambda)k^{+(l)}_{33}(\lambda) + 2k^{-(m)}_{11}(\lambda)k^{-(m)}_{33}(\lambda)
k^{+(l)}_{22}(\lambda)k^{+(l)}_{33}(\lambda)
\nonumber \\
&-& 2(k^{-(m)}_{33}(\lambda))^{2}k^{+(l)}_{22}(\lambda)k^{+(l)}_{33}(\lambda) +
k^{-(m)}_{11}(\lambda)k^{-(m)}_{22}(\lambda)(k^{+(l)}_{33}(\lambda))^{2}
\nonumber \\
&+& 2k^{-(m)}_{23}(\lambda)k^{-(m)}_{32}(\lambda)(k^{+(l)}_{33}(\lambda))^{2} +
k^{-(m)}_{11}(\lambda)k^{-(m)}_{33}(\lambda)(k^{+(l)}_{33}(\lambda))^{2}
\nonumber \\
&-& 2k^{-(m)}_{22}(\lambda)k^{-(m)}_{33}(\lambda)(k^{+(l)}_{33}(\lambda))^{2}
\end{eqnarray}
\begin{eqnarray}
Z^{(l,m)}_{5} &=& (k^{-(m)}_{11}(\lambda))^{2}(k^{+(l)}_{22}(\lambda))^{2} - k^{-(m)}_{11}(\lambda)k^{-(m)}_{22}(\lambda)(k^{+(l)}_{22}(\lambda))^{2} -
k^{-(m)}_{23}(\lambda)k^{-(m)}_{32}(\lambda)(k^{+(l)}_{22}(\lambda))^{2}
\nonumber \\
&-& k^{-(m)}_{11}(\lambda)k^{-(m)}_{33}(\lambda)(k^{+(l)}_{22}(\lambda))^{2} +
k^{-(m)}_{22}(\lambda)k^{-(m)}_{33}(\lambda)(k^{+(l)}_{22}(\lambda))^{2}
\nonumber \\
&-& (k^{-(m)}_{22}(\lambda))^{2}k^{+(l)}_{23}(\lambda)k^{+(l)}_{32}(\lambda) -
4k^{-(m)}_{23}(\lambda)k^{-(m)}_{32}(\lambda)k^{+(l)}_{23}(\lambda)k^{+(l)}_{32}(\lambda)
\nonumber \\
&+& 2k^{-(m)}_{22}(\lambda)k^{-(m)}_{33}(\lambda)
k^{+(l)}_{23}(\lambda)k^{+(l)}_{32}(\lambda) - (k^{-(m)}_{33}(\lambda))^{2}k^{+(l)}_{23}(\lambda)k^{+(l)}_{32}(\lambda)
\nonumber \\
&+& 2(k^{-(m)}_{11}(\lambda))^{2}k^{+(l)}_{22}(\lambda)k^{+(l)}_{33}(\lambda) - 2k^{-(m)}_{11}(\lambda)k^{-(m)}_{22}(\lambda)k^{+(l)}_{22}(\lambda)
k^{+(l)}_{33}(\lambda)
\nonumber \\
&+& (k^{-(m)}_{22}(\lambda))^{2}k^{+(l)}_{22}(\lambda)k^{+(l)}_{33}(\lambda) +
2k^{-(m)}_{23}(\lambda)k^{-(m)}_{32}(\lambda)k^{+(l)}_{22}(\lambda)k^{+(l)}_{33}(\lambda)
\nonumber \\
&-& 2k^{-(m)}_{11}(\lambda)k^{-(m)}_{33}(\lambda)
k^{+(l)}_{22}(\lambda)k^{+(l)}_{33}(\lambda) + (k^{-(m)}_{33}(\lambda))^{2}k^{+(l)}_{22}(\lambda)k^{+(l)}_{33}(\lambda)
\nonumber \\
&+& (k^{-(m)}_{11}(\lambda))^{2}(k^{+(l)}_{33}(\lambda))^{2} - k^{-(m)}_{11}(\lambda)k^{-(m)}_{22}(\lambda)(k^{+(l)}_{33}(\lambda))^{2} -
k^{-(m)}_{23}(\lambda)k^{-(m)}_{32}(\lambda)(k^{+(l)}_{33}(\lambda))^{2}
\nonumber \\
&-& k^{-(m)}_{11}(\lambda)k^{-(m)}_{33}(\lambda)(k^{+(l)}_{33}(\lambda))^{2} +
k^{-(m)}_{22}(\lambda)k^{-(m)}_{33}(\lambda)(k^{+(l)}_{33}(\lambda))^{2}
\end{eqnarray}
\begin{eqnarray}
Z^{(l,m)}_{4} &=& (k^{-(m)}_{11}(\lambda))^{2}(k^{+(l)}_{22}(\lambda))^{2} - 2k^{-(m)}_{11}(\lambda)k^{-(m)}_{22}(\lambda)(k^{+(l)}_{22}(\lambda))^{2} -
3(k^{-(m)}_{22}(\lambda))^{2}(k^{+(l)}_{22}(\lambda))^{2}
\nonumber \\
&-& 4k^{-(m)}_{23}(\lambda)k^{-(m)}_{32}(\lambda)(k^{+(l)}_{22}(\lambda))^{2} +
4k^{-(m)}_{22}(\lambda)k^{-(m)}_{33}(\lambda)(k^{+(l)}_{22}(\lambda))^{2}
\nonumber \\
&-& 2k^{-(m)}_{11}(\lambda)k^{-(m)}_{32}(\lambda)k^{+(l)}_{22}(\lambda)
k^{+(l)}_{23}(\lambda) - 6k^{-(m)}_{22}(\lambda)k^{-(m)}_{32}(\lambda)k^{+(l)}_{22}(\lambda)k^{+(l)}_{23}(\lambda)
\nonumber \\
&-& 3(k^{-(m)}_{32}(\lambda))^{2}(k^{+(l)}_{23}(\lambda))^{2} - 2k^{-(m)}_{11}(\lambda)k^{-(m)}_{23}(\lambda)k^{+(l)}_{22}(\lambda)k^{+(l)}_{32}(\lambda)
\nonumber \\
&-& 6k^{-(m)}_{22}(\lambda)k^{-(m)}_{23}(\lambda)k^{+(l)}_{22}(\lambda)k^{+(l)}_{32}(\lambda) - 4(k^{-(m)}_{22}(\lambda))^{2}k^{+(l)}_{23}(\lambda)
k^{+(l)}_{32}(\lambda)
\nonumber \\
&-& 6k^{-(m)}_{23}(\lambda)k^{-(m)}_{32}(\lambda)k^{+(l)}_{23}(\lambda)k^{+(l)}_{32}(\lambda) -
8k^{-(m)}_{22}(\lambda)k^{-(m)}_{33}(\lambda)k^{+(l)}_{23}(\lambda)k^{+(l)}_{32}(\lambda)
\nonumber \\
&-& 4(k^{-(m)}_{33}(\lambda))^{2}k^{+(l)}_{23}(\lambda)
k^{+(l)}_{32}(\lambda) - 3(k^{-(m)}_{23}(\lambda))^{2}(k^{+(l)}_{32}(\lambda))^{2}
\nonumber \\
&+& 2(k^{-(m)}_{11}(\lambda))^{2}k^{+(l)}_{22}(\lambda)
k^{+(l)}_{33}(\lambda) - 2k^{-(m)}_{11}(\lambda)k^{-(m)}_{22}(\lambda)k^{+(l)}_{22}(\lambda)k^{+(l)}_{33}(\lambda)
\nonumber \\
&+& 4(k^{-(m)}_{22}(\lambda))^{2}k^{+(l)}_{22}(\lambda)k^{+(l)}_{33}(\lambda) - 8k^{-(m)}_{23}(\lambda)k^{-(m)}_{32}(\lambda)k^{+(l)}_{22}(\lambda)
k^{+(l)}_{33}(\lambda)
\nonumber \\
&-& 2k^{-(m)}_{11}(\lambda)k^{-(m)}_{33}(\lambda)k^{+(l)}_{22}(\lambda)k^{+(l)}_{33}(\lambda) +
10k^{-(m)}_{22}(\lambda)k^{-(m)}_{33}(\lambda)k^{+(l)}_{22}(\lambda)k^{+(l)}_{33}(\lambda)
\nonumber \\
&+& 4(k^{-(m)}_{33}(\lambda))^{2}k^{+(l)}_{22}(\lambda)
k^{+(l)}_{33}(\lambda)
- 2k^{-(m)}_{11}(\lambda)k^{-(m)}_{32}(\lambda)k^{+(l)}_{23}(\lambda)k^{+(l)}_{33}(\lambda)
\nonumber \\
&-& 6k^{-(m)}_{32}(\lambda)k^{-(m)}_{33}(\lambda)k^{+(l)}_{23}(\lambda)k^{+(l)}_{33}(\lambda) - 2k^{-(m)}_{11}(\lambda)k^{-(m)}_{23}(\lambda)
k^{+(l)}_{32}(\lambda)k^{+(l)}_{33}(\lambda)
\nonumber \\
&-& 6k^{-(m)}_{23}(\lambda)k^{-(m)}_{33}(\lambda)k^{+(l)}_{32}(\lambda)k^{+(l)}_{33}(\lambda) +
(k^{-(m)}_{11}(\lambda))^{2}(k^{+(l)}_{33}(\lambda))^{2}
\nonumber \\
&-& 4k^{-(m)}_{23}(\lambda)k^{-(m)}_{32}(\lambda)(k^{+(l)}_{33}(\lambda))^{2} -
2k^{-(m)}_{11}(\lambda)k^{-(m)}_{33}(\lambda)(k^{+(l)}_{33}(\lambda))^{2}
\nonumber \\
&+& 4k^{-(m)}_{22}(\lambda)k^{-(m)}_{33}(\lambda)(k^{+(l)}_{33}(\lambda))^{2} -
3(k^{-(m)}_{33}(\lambda))^{2}(k^{+(l)}_{33}(\lambda))^{2}
\end{eqnarray}
\begin{eqnarray}
Z^{(l,m)}_{3} &=& k^{-(m)}_{11}(\lambda)k^{-(m)}_{22}(\lambda)(k^{+(l)}_{22}(\lambda))^{2} - 2(k^{-(m)}_{22}(\lambda))^{2}(k^{+(l)}_{22}(\lambda))^{2} -
k^{-(m)}_{11}(\lambda)k^{-(m)}_{33}(\lambda)(k^{+(l)}_{22}(\lambda))^{2}
\nonumber \\
&+& 2k^{-(m)}_{11}(\lambda)k^{-(m)}_{32}(\lambda)k^{+(l)}_{22}(\lambda)
k^{+(l)}_{23}(\lambda) - 4k^{-(m)}_{22}(\lambda)k^{-(m)}_{32}(\lambda)k^{+(l)}_{22}(\lambda)k^{+(l)}_{23}(\lambda)
\nonumber \\
&-& 2(k^{-(m)}_{32}(\lambda))^{2}(k^{+(l)}_{23}(\lambda))^{2} + 2k^{-(m)}_{11}(\lambda)k^{-(m)}_{23}(\lambda)k^{+(l)}_{22}(\lambda)k^{+(l)}_{32}(\lambda)
\nonumber \\
&-& 4k^{-(m)}_{22}(\lambda)k^{-(m)}_{23}(\lambda)k^{+(l)}_{22}(\lambda)k^{+(l)}_{32}(\lambda) + 4k^{-(m)}_{11}(\lambda)k^{-(m)}_{22}(\lambda)
k^{+(l)}_{23}(\lambda)k^{+(l)}_{32}(\lambda)
\nonumber \\
&-& 4(k^{-(m)}_{22}(\lambda))^{2}k^{+(l)}_{23}(\lambda)k^{+(l)}_{32}(\lambda) -
4k^{-(m)}_{23}(\lambda)k^{-(m)}_{32}(\lambda)k^{+(l)}_{23}(\lambda)k^{+(l)}_{32}(\lambda)
\nonumber \\
&+& 4k^{-(m)}_{11}(\lambda)k^{-(m)}_{33}(\lambda)
k^{+(l)}_{23}(\lambda)k^{+(l)}_{32}(\lambda) - 8k^{-(m)}_{22}(\lambda)k^{-(m)}_{33}(\lambda)k^{+(l)}_{23}(\lambda)k^{+(l)}_{32}(\lambda)
\nonumber \\
&-& 4(k^{-(m)}_{33}(\lambda))^{2}k^{+(l)}_{23}(\lambda)k^{+(l)}_{32}(\lambda) - 2(k^{-(m)}_{23}(\lambda))^{2}(k^{+(l)}_{32}(\lambda))^{2}
\nonumber \\
&-& 4k^{-(m)}_{11}(\lambda)k^{-(m)}_{22}(\lambda)k^{+(l)}_{22}(\lambda)k^{+(l)}_{33}(\lambda) + 4(k^{-(m)}_{22}(\lambda))^{2}k^{+(l)}_{22}(\lambda)
k^{+(l)}_{33}(\lambda)
\nonumber \\
&-& 4k^{-(m)}_{11}(\lambda)k^{-(m)}_{33}(\lambda)k^{+(l)}_{22}(\lambda)k^{+(l)}_{33}(\lambda) +
4k^{-(m)}_{22}(\lambda)k^{-(m)}_{33}(\lambda)k^{+(l)}_{22}(\lambda)k^{+(l)}_{33}(\lambda)
\nonumber \\
&+& 4(k^{-(m)}_{33}(\lambda))^{2}k^{+(l)}_{22}(\lambda)
k^{+(l)}_{33}(\lambda) + 2k^{-(m)}_{11}(\lambda)k^{-(m)}_{32}(\lambda)k^{+(l)}_{23}(\lambda)k^{+(l)}_{33}(\lambda)
\nonumber \\
&-& 4k^{-(m)}_{32}(\lambda)k^{-(m)}_{33}(\lambda)k^{+(l)}_{23}(\lambda)k^{+(l)}_{33}(\lambda) + 2k^{-(m)}_{11}(\lambda)k^{-(m)}_{23}(\lambda)
k^{+(l)}_{32}(\lambda)k^{+(l)}_{33}(\lambda)
\nonumber \\
&-& 4k^{-(m)}_{23}(\lambda)k^{-(m)}_{33}(\lambda)k^{+(l)}_{32}(\lambda)k^{+(l)}_{33}(\lambda) -
k^{-(m)}_{11}(\lambda)k^{-(m)}_{22}(\lambda)(k^{+(l)}_{33}(\lambda))^{2}
\nonumber \\
&+& k^{-(m)}_{11}(\lambda)k^{-(m)}_{33}(\lambda)(k^{+(l)}_{33}(\lambda))^{2} -
2(k^{-(m)}_{33}(\lambda))^{2}(k^{+(l)}_{33}(\lambda))^{2}
\end{eqnarray}
\begin{eqnarray}
Z^{(l,m)}_{2} &=& 3k^{-(m)}_{11}(\lambda)k^{-(m)}_{22}(\lambda)(k^{+(l)}_{22}(\lambda))^2 - (k^{-(m)}_{22}(\lambda))^2(k^{+(l)}_{22}(\lambda))^2 +
3k^{-(m)}_{23}(\lambda)k^{-(m)}_{32}(\lambda)(k^{+(l)}_{22}(\lambda))^2
\nonumber \\
&+& k^{-(m)}_{11}(\lambda)k^{-(m)}_{33}(\lambda)(k^{+(l)}_{22}(\lambda))^2 -
3k^{-(m)}_{22}(\lambda)k^{-(m)}_{33}(\lambda)(k^{+(l)}_{22}(\lambda))^2
\nonumber \\
&+& 2k^{-(m)}_{11}(\lambda)k^{-(m)}_{32}(\lambda)k^{+(l)}_{22}(\lambda)
k^{+(l)}_{23}(\lambda) - 2k^{-(m)}_{22}(\lambda)k^{-(m)}_{32}(\lambda)k^{+(l)}_{22}(\lambda)k^{+(l)}_{23}(\lambda)
\nonumber \\
&-& (k^{-(m)}_{32}(\lambda))^2(k^{+(l)}_{23}(\lambda))^2 + 2k^{-(m)}_{11}(\lambda)k^{-(m)}_{23}(\lambda)k^{+(l)}_{22}(\lambda)k^{+(l)}_{32}(\lambda)
\nonumber \\
&-& 2k^{-(m)}_{22}(\lambda)k^{-(m)}_{23}(\lambda)k^{+(l)}_{22}(\lambda)k^{+(l)}_{32}(\lambda) - 4(k^{-(m)}_{11}(\lambda))^2k^{+(l)}_{23}(\lambda)
k^{+(l)}_{32}(\lambda)
\nonumber \\
&+& 12k^{-(m)}_{11}(\lambda)k^{-(m)}_{22}(\lambda)k^{+(l)}_{23}(\lambda)k^{+(l)}_{32}(\lambda) -
5(k^{-(m)}_{22}(\lambda))^2k^{+(l)}_{23}(\lambda)k^{+(l)}_{32}(\lambda)
\nonumber \\
&+& 2k^{-(m)}_{23}(\lambda)k^{-(m)}_{32}(\lambda)k^{+(l)}_{23}(\lambda)
k^{+(l)}_{32}(\lambda) + 12k^{-(m)}_{11}(\lambda)k^{-(m)}_{33}(\lambda)k^{+(l)}_{23}(\lambda)k^{+(l)}_{32}(\lambda)
\nonumber \\
&-& 14k^{-(m)}_{22}(\lambda)k^{-(m)}_{33}(\lambda)k^{+(l)}_{23}(\lambda)k^{+(l)}_{32}(\lambda) - 5(k^{-(m)}_{33}(\lambda))^2k^{+(l)}_{23}(\lambda)
k^{+(l)}_{32}(\lambda)
\nonumber \\
&-& (k^{-(m)}_{23}(\lambda))^2(k^{+(l)}_{32}(\lambda))^2 + 4(k^{-(m)}_{11}(\lambda))^2k^{+(l)}_{22}(\lambda)
k^{+(l)}_{33}(\lambda) - (k^{-(m)}_{33}(\lambda))^2(k^{+(l)}_{33}(\lambda))^2
\nonumber \\
&-& 8k^{-(m)}_{11}(\lambda)k^{-(m)}_{22}(\lambda)k^{+(l)}_{22}(\lambda)k^{+(l)}_{33}(\lambda) +
5(k^{-(m)}_{22}(\lambda))^2k^{+(l)}_{22}(\lambda)k^{+(l)}_{33}(\lambda)
\nonumber \\
&+& 2k^{-(m)}_{23}(\lambda)k^{-(m)}_{32}(\lambda)k^{+(l)}_{22}(\lambda)
k^{+(l)}_{33}(\lambda) - 8k^{-(m)}_{11}(\lambda)k^{-(m)}_{33}(\lambda)k^{+(l)}_{22}(\lambda)k^{+(l)}_{33}(\lambda)
\nonumber \\
&+& 6k^{-(m)}_{22}(\lambda)k^{-(m)}_{33}(\lambda)k^{+(l)}_{22}(\lambda)k^{+(l)}_{33}(\lambda) + 5(k^{-(m)}_{33}(\lambda))^2k^{+(l)}_{22}(\lambda)
k^{+(l)}_{33}(\lambda)
\nonumber \\
&+& 2k^{-(m)}_{11}(\lambda)k^{-(m)}_{32}(\lambda)k^{+(l)}_{23}(\lambda)k^{+(l)}_{33}(\lambda) -
2k^{-(m)}_{32}(\lambda)k^{-(m)}_{33}(\lambda)k^{+(l)}_{23}(\lambda)k^{+(l)}_{33}(\lambda)
\nonumber \\
&+& 2k^{-(m)}_{11}(\lambda)k^{-(m)}_{23}(\lambda)
k^{+(l)}_{32}(\lambda)k^{+(l)}_{33}(\lambda) - 2k^{-(m)}_{23}(\lambda)k^{-(m)}_{33}(\lambda)k^{+(l)}_{32}(\lambda)k^{+(l)}_{33}(\lambda)
\nonumber \\
&+& k^{-(m)}_{11}(\lambda)k^{-(m)}_{22}(\lambda)(k^{+(l)}_{33}(\lambda))^2 + 3k^{-(m)}_{23}(\lambda)k^{-(m)}_{32}(\lambda)(k^{+(l)}_{33}(\lambda))^2
\nonumber \\
&+& 3k^{-(m)}_{11}(\lambda)k^{-(m)}_{33}(\lambda)(k^{+(l)}_{33}(\lambda))^2 - 3k^{-(m)}_{22}(\lambda)k^{-(m)}_{33}(\lambda)(k^{+(l)}_{33}(\lambda))^2
\end{eqnarray}
\begin{eqnarray}
Z^{(l,m)}_{1} &=& (k^{-(m)}_{11}(\lambda))^{2}(k^{+(l)}_{22}(\lambda))^{2} - k^{-(m)}_{11}(\lambda)k^{-(m)}_{22}(\lambda)(k^{+(l)}_{22}(\lambda))^{2} -
k^{-(m)}_{23}(\lambda)k^{-(m)}_{32}(\lambda)(k^{+(l)}_{22}(\lambda))^{2}
\nonumber \\
&-& k^{-(m)}_{11}(\lambda)k^{-(m)}_{33}(\lambda)(k^{+(l)}_{22}(\lambda))^{2} +
k^{-(m)}_{22}(\lambda)k^{-(m)}_{33}(\lambda)(k^{+(l)}_{22}(\lambda))^{2}
\nonumber \\
&+& 4(k^{-(m)}_{11}(\lambda))^{2}k^{+(l)}_{23}(\lambda)k^{+(l)}_{32}(\lambda) -
6k^{-(m)}_{11}(\lambda)k^{-(m)}_{22}(\lambda)k^{+(l)}_{23}(\lambda)k^{+(l)}_{32}(\lambda)
\nonumber \\
&+& (k^{-(m)}_{22}(\lambda))^{2}k^{+(l)}_{23}(\lambda)
k^{+(l)}_{32}(\lambda) - 4k^{-(m)}_{23}(\lambda)k^{-(m)}_{32}(\lambda)k^{+(l)}_{23}(\lambda)k^{+(l)}_{32}(\lambda)
\nonumber \\
&-& 6k^{-(m)}_{11}(\lambda)k^{-(m)}_{33}(\lambda)k^{+(l)}_{23}(\lambda)k^{+(l)}_{32}(\lambda) + 6k^{-(m)}_{22}(\lambda)k^{-(m)}_{33}(\lambda)
k^{+(l)}_{23}(\lambda)k^{+(l)}_{32}(\lambda)
\nonumber \\
&+& (k^{-(m)}_{33}(\lambda))^{2}k^{+(l)}_{23}(\lambda)k^{+(l)}_{32}(\lambda) -
2(k^{-(m)}_{11}(\lambda))^{2}k^{+(l)}_{22}(\lambda)k^{+(l)}_{33}(\lambda)
\nonumber \\
&+& 4k^{-(m)}_{11}(\lambda)k^{-(m)}_{22}(\lambda)k^{+(l)}_{22}(\lambda)
k^{+(l)}_{33}(\lambda) - (k^{-(m)}_{22}(\lambda))^{2}k^{+(l)}_{22}(\lambda)k^{+(l)}_{33}(\lambda)
\nonumber \\
&+& 2k^{-(m)}_{23}(\lambda)k^{-(m)}_{32}(\lambda)k^{+(l)}_{22}(\lambda)k^{+(l)}_{33}(\lambda) + 4k^{-(m)}_{11}(\lambda)k^{-(m)}_{33}(\lambda)
k^{+(l)}_{22}(\lambda)k^{+(l)}_{33}(\lambda)
\nonumber \\
&-& 4k^{-(m)}_{22}(\lambda)k^{-(m)}_{33}(\lambda)k^{+(l)}_{22}(\lambda)k^{+(l)}_{33}(\lambda) -
(k^{-(m)}_{33}(\lambda))^{2}k^{+(l)}_{22}(\lambda)k^{+(l)}_{33}(\lambda)
\nonumber \\
&+& (k^{-(m)}_{11}(\lambda))^{2}(k^{+(l)}_{33}(\lambda))^{2} -
k^{-(m)}_{11}(\lambda)k^{-(m)}_{22}(\lambda)(k^{+(l)}_{33}(\lambda))^{2} - k^{-(m)}_{23}(\lambda)k^{-(m)}_{32}(\lambda)(k^{+(l)}_{33}(\lambda))^{2}
\nonumber \\
&-& k^{-(m)}_{11}(\lambda)k^{-(m)}_{33}(\lambda)(k^{+(l)}_{33}(\lambda))^{2} + k^{-(m)}_{22}(\lambda)k^{-(m)}_{33}(\lambda)(k^{+(l)}_{33}(\lambda))^{2}
\end{eqnarray}
\begin{eqnarray}
Z^{(l,m)}_{0} &=& ((k^{-(m)}_{11}(\lambda))^{2} - k^{-(m)}_{11}(\lambda)k^{-(m)}_{22}(\lambda) - k^{-(m)}_{23}(\lambda)k^{-(m)}_{32}(\lambda) -
k^{-(m)}_{11}(\lambda)k^{-(m)}_{33}(\lambda)
\nonumber \\
&+& k^{-(m)}_{22}(\lambda)k^{-(m)}_{33}(\lambda))(k^{+(l)}_{22}(\lambda)k^{+(l)}_{33}(\lambda) -
k^{+(l)}_{23}(\lambda)k^{+(l)}_{32}(\lambda))
\end{eqnarray}


\begin{thebibliography}{}
\bibitem{ZR} F.C. Zhang and T.M. Rice, {\it Phys. Rev. B 37, 3759 (1988)}.
\bibitem{SCH1} P. Schlottmann, {\it Physica B 171, 30 (1991)}.
\bibitem{SCH2} P. Schlottmann, {\it Phys. Rev. B 36, 5177 (1987)}.
\bibitem{GROS} C. Gros, R. Joynt and T.M. Rice, {\it Phys. Rev. B 36, 381 (1987)}.
\bibitem{KANE} C.L. Kane, P.A. Lee and N. Read, {\it Phys. Rev. B 39, 6880 (1989)}.
\bibitem{AND} P.W. Anderson, {\it Science 235, 1196 (1987)}.
\bibitem{AND1} P.W. Anderson, {\it Phys. Rev. Lett. 64, 1839 (1990)};
P.W. Anderson, {\it Phys. Rev. Lett. 65, 2306 (1990)}.
\bibitem{BRINK} W.F. Brinkman and T.M. Rice, {\it Phys. Rev. B 2, 1324 (1970)}.
\bibitem{SCH3} P. Schlottmann, {\it Int. J. Mod. Phys. B 11, 355 (1997)}.
\bibitem{LAI} C.K. Lai, {\it J. Math. Phys. 15, 1675 (1974)}.
\bibitem{SU} B.Sutherland, {\it Phys. Rev. B 12, 3795 (1975)}.
\bibitem{SAR} S. Sarkar, {\it J. Phys. A: Math. Gen 23, L409 (1990)}.
\bibitem{BAR} P.A. Bares and G. Blatter, {\it Phys. Rev. Lett. 64, 2567 (1990)};
P.A. Bares, G. Blatter and M. Ogata, {\it Phys. Rev. B 44, 130 (1991)}.
\bibitem{KOR} F.H.L. Essler and V.E. Korepin, {\it Phys. Rev. B 46, 9147 (1992)}.
\bibitem{CARDY} J.L. Cardy, {\it Nucl. Phys. B 275, 200 (1986)}.
\bibitem{ALCARAZ} F.C. Alcaraz, M.N. Barber and M.T. Batchelor, {\it Ann. Phys. 182, 280 (1988)}.
\bibitem{SK} E.K. Skylianin, {\it J. Phys. A: Math. Gen. 21, 2375 (1988)}.
\bibitem{FAD} L.D. Takhtajan and L.D. Faddeev, {\it Russian Math. Surveys 34 (1979) 11};
V.E. Korepin, G. Izergin and N.M. Bogoliubov, {\it ``Quantum Inverse Scattering Method and
Correlation Functions'', Cambridge Univ. Press, Cambridge, (1993)}.
\bibitem{CHER} I.V. Cherednik, {\it Theor. Math. Phys. 61, 977 (1984)}.
\bibitem{GONZA} A. Gonz\'alez-Ruiz, {\it Nucl. Phys. B 424, 468 (1994)}.
\bibitem{ESSLER} F.H.L. Essler, {\it J. Phys. A: Math. Gen 29, 6183 (1996)}.
\bibitem{SASAKI} W.L. Yang and R. Sasaki, {\it Nucl. Phys. B 679, 495 (2004)}.
\bibitem{WG} W. Galleas and M.J. Martins, {\it Phys. Lett. A 335, 167 (2005)}.
\bibitem{CLA} C.S Melo, M.J. Martins and G.A.P. Ribeiro, {\it Nucl. Phys. B 711, 565 (2005)}.
\bibitem{FAN} H. Fan, B.Y. Hou, K.J. Shi and Z.X. Yang, {\it Nucl. Phys. B 478, 723 (1996)};
W.L. Yang and Y.Z. Zhang, {\it Nucl. Phys. B 744, 312 (2006)}.
\bibitem{NEPO} R. Murgan, R.I. Nepomechie and C. Shi, {\it J. Stat. Mech.: Theor. Exp., P08006 (2004)};
W.L. Yang, R.I. Nepomechie and  Y.Z. Zhang, {\it  Phys. Lett. B 633, 664 (2006)};
R. Murgan and  R.I. Nepomechie, {\it J. Stat. Mech.: Theor. Exp., P08002 (2005)};
R. Murgan and  R.I. Nepomechie, {\it J. Stat. Mech.: Theor. Exp., P05007 (2005)};
R.I. Nepomechie, {\it  J. Phys. A: Math. Gen 37, 433 (2004)}.
\bibitem{CAO} J.P. Cao, H.Q. Lin, K.J. Shi and Y.P. Wang, {\it Nucl. Phys. B 663, 487 (2003)}.
\bibitem{GIER} J. de Gier and P. Pyatov, {\it J. Stat. Mech.: Theor. Exp., P002 (2004)}.
\bibitem{ABAD} J. Abad and M. Rios, {\it Phys. Lett. B 352 , 92 (1995)}.
\bibitem{LIMA} R. Malara and A. Lima-Santos, {\it J. Stat. Mech.: Theor. Exp., P09013 (2006)}.
\bibitem{LM} A. Lima-Santos and M.J. Martins, {\it Nucl. Phys. B 760 ,184 (2007)}.
\bibitem{ALS} A. Lima-Santos, {\it Nucl. Phys. B 558, 637 (1999)}.
\bibitem{GUANG0} A.J. Bracken, X.Y. Ge, Y.Z. Zhang and H.Q. Zhou, {\it Nucl. Phys. B 516, 588 (1998)};
M.J. Martins and X.W. Guan, {\it Nucl. Phys. B 562, 433 (1999)};
G.L. Li, R.H. Yue and B.Y. Hou, {\it Nucl. Phys. B 586, 711 (2000)};
G.L. Li, K.J. Shi and R.H. Yue, {\it Nucl. Phys. B 687, 220 (2004)}.
\bibitem{GUANG1} G.L. Li and K.J. Shi, {\it J. Stat. Mech.: Theor. Exp., P01018 (2007)}.
\bibitem{B11} L. Mezincescu and R.I Nepomechie, {\it J. Phys. A: Math. Gen 24, 17 (1991)};
L. Mezincescu and R.I Nepomechie, {\it Int. J. Mod. Phys. A 6, 5231 (1991)}.
\bibitem{B12} H.J. de Vega and A. Gonz\'alez-Ruiz, {\it Mod. Phys. Lett. A 9, 2207 (1994)}.
\bibitem{PPP} A. Foerster and M. Karowski, {\it Nucl. Phys. B 408, 512 (1993)};
H.J. de Vega and A. Gonz\'alez-Ruiz, {\it Nucl. Phys. B 417, 553 (1994)}.
\end{thebibliography}
\end{document}